\documentclass[aps,pra,reprint, amsmath, amssymb,superscriptaddress,longbibliography]{revtex4-1}
\usepackage{bm}
\usepackage[retainorgcmds]{IEEEtrantools}
\usepackage{graphicx}
\usepackage{mathrsfs}
\usepackage{amsmath}
\usepackage{amsfonts}
\usepackage{amssymb}
\usepackage{color}
\usepackage{times,txfonts}
\usepackage{nicefrac}
\usepackage[normalem]{ulem}
\usepackage{ragged2e}
\usepackage{tikz}

\usepackage[colorlinks=true,linkcolor=blue,urlcolor=blue,citecolor=blue,pdfusetitle]{hyperref}

\usepackage{blindtext}

\newcommand{\ket}[1]{\left|#1\right\rangle}
\newcommand{\bra}[1]{\langle#1|}

\begin{document}

\title{Optimal Control in Disordered Quantum Systems}

\date{\today}
\author{Luuk Coopmans}
\email{luuk.coopmans@cambridgequantum.com}
\affiliation{Dublin Institute for Advanced Studies, School of Theoretical Physics, 10 Burlington Rd, Dublin, Ireland}
\affiliation{School of Physics, Trinity College Dublin, College Green, Dublin 2, Ireland}
\affiliation{Quantinuum, Partnership House, Carlisle Place, London SW1P 1BX, United Kingdom}
\author{Steve Campbell}
\affiliation{School of Physics, University College Dublin, Belfield, Dublin 4, Ireland}
\affiliation{Centre for Quantum Engineering, Science, and Technology, University College Dublin, Belfield, Dublin 4, Ireland}
\author{Gabriele De Chiara}
\affiliation{Centre for Quantum Materials and Technology, School of Mathematics and Physics,
Queen’s University Belfast, Belfast BT7 1NN, United Kingdom}
\author{Anthony Kiely}
\affiliation{School of Physics, University College Dublin, Belfield, Dublin 4, Ireland}
\affiliation{Centre for Quantum Engineering, Science, and Technology, University College Dublin, Belfield, Dublin 4, Ireland}

\begin{abstract}
We investigate several control strategies for the transport of an excitation along a spin chain. We demonstrate that fast, high fidelity transport can be achieved using protocols designed with differentiable programming. Building on this, we then show how this approach can be effectively adapted to control a disordered quantum system. We consider two settings: optimal control for a known unwanted disorder pattern, i.e. a specific disorder realisation, and optimal control where only the statistical properties of disorder are known, i.e. optimizing for high average fidelities. In the former, disorder effects can be effectively mitigated for an appropriately chosen control protocol. However, in the latter setting the average fidelity can only be marginally improved, suggesting the presence of a fundamental lower bound.
\end{abstract}
\maketitle{}

\section{Introduction \label{intro}}
Robust, implementable control protocols are a necessary ingredient for many quantum devices. A particularly relevant task is the routing of quantum states and information over large distances~\cite{Zoller2005, Kimble2008} and/or across a network of connected quantum registers~\cite{Majer2007, BarisPRA}. The use of spin chains with short range Heisenberg interactions to transmit quantum information presents a viable framework to accomplish this goal~\cite{Bose2003}, applicable to a wide range of experimental platforms~\cite{Demler2003, Romito2005, Cappellaro2007, Hild2014, Qiao2020}. This task can be achieved via the use of an external magnetic field with a parabolic spatial profile which is adiabatically swept across the chain~\cite{Vbalachandran2008} alleviating the need for precise control over individual couplings~\cite{Eckert2007}. However, a notable drawback is that such adiabatic protocols are inherently slow and hence susceptible to noise from the environment. This observation precipitated the development of non-adiabatic protocols such as analytically derived shortcuts to adiabaticity (STA)~\cite{Kiely2021} and quantum optimal control methods~\cite{GrossJPB, glaser2015, Koch2022, Muller2021, GabrielePRB, Bukov, Referee6, Referee7, Referee8, Referee9, Referee10, Referee11, Referee12, Referee13}.

While these approaches have been successfully applied in several settings, there remain drawbacks, for instance shortcut-to-adiabaticity approaches, such as the one employed in Ref.~\cite{Kiely2021}, often rely on simple model descriptions and/or approximations. In this work we augment traditional control techniques with differentiable programming ($\partial$P)~\cite{Wengert1964, Liao2019} and critically examine their efficacy. Differential programming is an approach which is widely used in the context of deep learning~\cite{Baydin2017} and has been recently applied in various quantum control setups, such as the transport of Majorana zero modes~\cite{Coopmans2021}, the control of thermal machines~\cite{khait2021optimal}, numerical renormalisation group~\cite{JonasPRR}, and interacting qubits~\cite{Schafer2020}. The main advantage of $\partial$P is that it allows to efficiently obtain the required gradients~\cite{Leung2017}, making gradient based optimization of complex quantum many-body problems computationally tractable.   

We use $\partial$P with several different ansatzes drawn from other traditional quantum control methods, specifically the shortcut-to-adiabaticity protocol~\cite{Kiely2021} and a finite Fourier basis inspired by the chopped random basis (CRAB) protocol~\cite{Muller2021}. A common difficulty which must be overcome is the inevitability of noise or imperfections in experimentally realised systems in form of spatial heterogeneities, i.e. disorder.  For sufficiently large disorder, the system's state tends to exponentially localize~\cite{Anderson1958} and thus hinders transport of the state across the system.  We establish that the presence of disorder is, in and of itself, not a limiting factor in achieving high fidelity control in the system, with $\partial$P strategies able to find a suitable control pulse efficiently. The drawback, however, is that the $\partial$P gradients implicitly depend on the exact disorder pattern, which therefore must be known. In the absence of this information, for a disordered-averaged gradient, we find that even with a high degree of optimization, there is little advantage gained over applying the best disorder-free protocols.

The remainder of the paper is organised as follows. In the subsequent section we outline the model and the state transfer goal. Then in Sec.~\ref{proto} we introduce the various approaches used to design the dynamical control field. Following this, we demonstrate and compare the effectiveness of these methods for a clean (i.e. disorder-free) system in Sec.~\ref{clean}. These methods are then adapted in Sec.~\ref{dirty} to a disordered system. Finally, in Sec.~\ref{conc} we summarize our work and provide some outlook on future directions.

\section{Model and Control Setup\label{model}}
We consider a one-dimensional Heisenberg model of $N$ interacting spin-$\frac{1}{2}$ particles. The Hamiltonian ($\hbar=1$) is 
\begin{equation}
\mathcal{H} = - \frac{J}{2} \sum_{i=1}^{N-1} \bm{\sigma}_{i}\cdot \bm{\sigma}_{i+1}+ \sum_{i=1}^{N} B_i \sigma_i^z ,
\label{eq:heis}
\end{equation} 
where $\bm{\sigma}_i\equiv\left(\sigma^x_i,\sigma^y_i, \sigma^z_i\right)$ are the Pauli matrices for the spin at site $i$. The constant isotropic nearest-neighbour coupling $J$ describes exchange interactions between the neighbouring spins and open boundary conditions are considered.

For a single excitation transport problem we do not need access to the full many-body Hilbert space as this Hamiltonian preserves the total spin excitation number, i.e. $S_z=\sum_{i=1}^N\sigma_i^z $, since $\left[\mathcal{H}, S_z \right]\!=\! 0$. We can divide the Hilbert space into separate disconnected sectors, each with a fixed total number of spin excitations $S\!=\!\langle S_z\rangle$. In what follows we focus on the single spin-excitation sector where one spin is up, denoted by $\ket{\uparrow}$, while all the other spins are down $\ket{\downarrow}$. In total we have $N$ such single-spin excitation states, which we label by $\ket{n}=\ket{\downarrow}_1\otimes\cdots\otimes\ket{\uparrow}_n\otimes\cdots\otimes\ket{\downarrow}_N$ where only the $n$-th spin is up. The Hamiltonian Eq.~\eqref{eq:heis} projected into this subspace becomes
\begin{align}
H =&\sum_{n=1}^N\left[B_n-2J\right]\ket{n}\bra{n} + J\ket{1}\bra{1}+ 
			J\ket{N}\bra{N} \nonumber \\ & + J\sum_{n=1}^{N-1} \ket{n}\bra{n+1} + \ket{n+1}\bra{n},
			\label{eq:seHam}
\end{align}
which is an exponential reduction in dimension compared to the full many-body Hamiltonian. 

Notice that there is a simple correspondence (up to some edge effects) between this single excitation Hamiltonian Eq.~\eqref{eq:seHam} and a discretized version of a single particle in a potential trap, whose continuum limit has the Hamiltonian 
\begin{equation}
H_c = \frac{k^2}{2m}+V(x). \label{eq:singlep}
\end{equation}

\begin{figure}[t]
    \centering
    \includegraphics[width=0.45\textwidth]{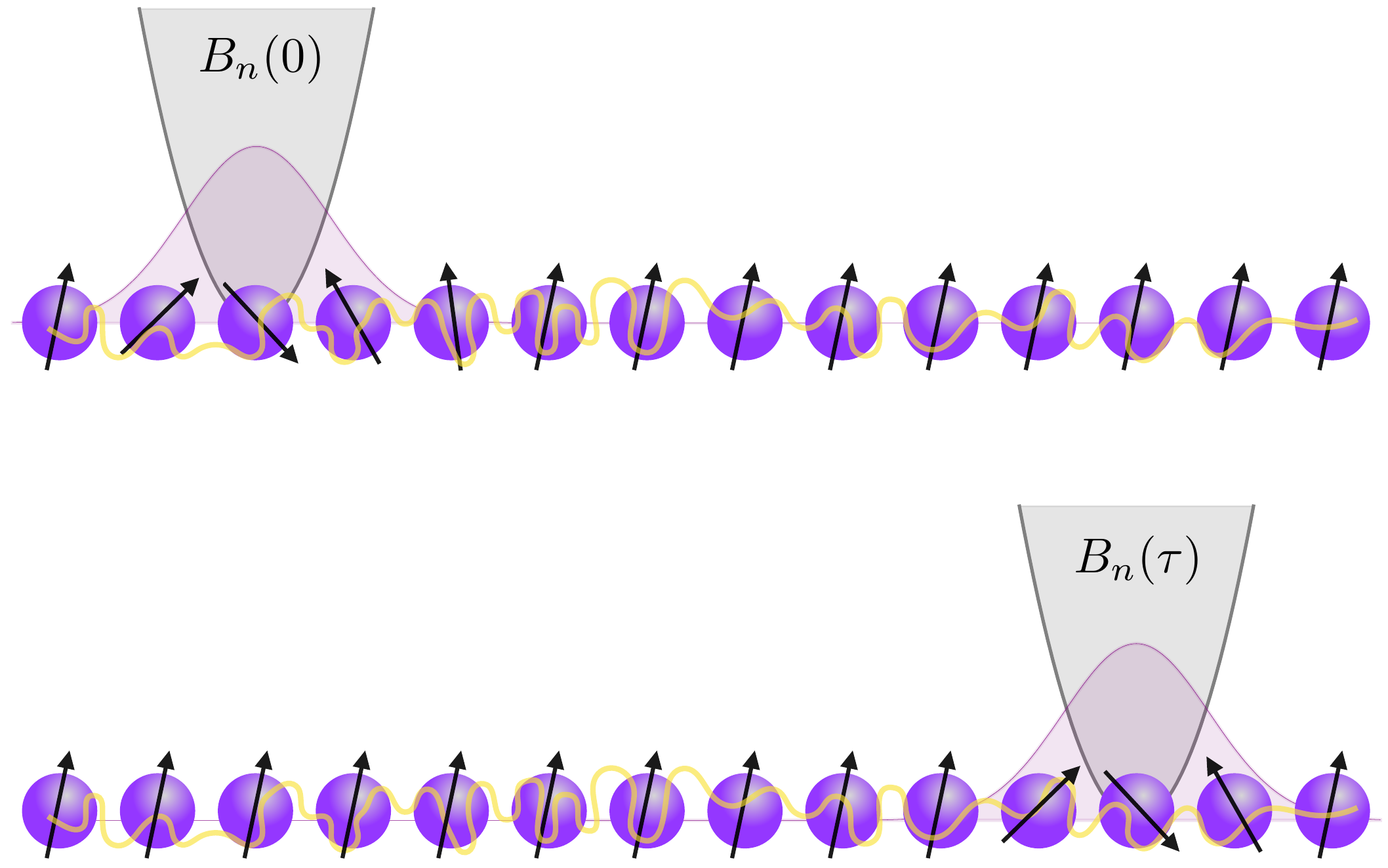}
    \caption{Setup of the disordered magnon transport problem. The control objective is to design the parabolic magnetic field $B_n$ (grey shading) in such a way that a localized wave packet or magnon excitation (purple shading) is transported across the chain. The disorder is uniform across the spin chain (yellow erratic line) and is static.}
    \label{fig:magnontransport}
\end{figure}

\subsection{Magnons and the Optimization Objective}
For our optimization problem we aim to transport magnons, single spin excitations, from one position in the chain to another, see Fig. \ref{fig:magnontransport}. Freely propagating magnons, i.e. without a confining magnetic field $B_n$, disperse over time throughout the chain. Therefore, an initially localized wave packet will delocalize across the chain as a direct result of the non-linear dispersion relation~\cite{Ahmed2017}. To counteract this spreading it is necessary to guide the magnon transport~\cite{Ahmed2015, Vbalachandran2008} by imposing the external magnetic field $B_n(t)$. Various kinds of magnetic traps with different spatial profiles can be used for this, such as, the P\"{o}schl-Teller~\cite{Makin2012} potential, a square well~\cite{Ahmed2017}, or a harmonic trap~\cite{Vbalachandran2008}. 

Motivated by the correspondence with the single-particle Hamiltonian in Eq.~\eqref{eq:seHam} we focus on the parabolic profile
\begin{equation}\label{eq:trap}
B_n(t) = -\frac{\omega^2}{4 J}\left[\frac{n-X_0(t)}{\Delta x}\right]^2\ket{n}\bra{n}.
\end{equation} 
$X_0(t)$  is   the position of the minimum of the trap, which will serve as our control parameter. This influences the position of the magnon. This functional form allows us to exploit known analytic optimal control pulses derived with STA methods~\cite{Torrontegui2011, Kiely2021}. Note that we fix the lattice spacing as $\Delta x=1$ hereafter.

For our magnon transport problem we initialize the parameters such that we have a Gaussian wave packet 
\begin{equation}\label{eq:initstate}
\ket{\psi_A} = \frac{1}{\sqrt{\sum_nc_n^2(x_A)}}\sum_nc_n(x_A)\ket{n}
\end{equation} 
centred at a position $x_A$ in the chain, with $c_n(x)\!\!=\!\!\exp{\left[-(n-x)^2/2\sigma^2\right]}$.  We vary $X_0(t)$ in such a way that we reach the target state 
\begin{equation}\label{eq:finstate}
\ket{\psi_B} = \frac{1}{\sqrt{\sum_nc_n^2(x_B)}}\sum_nc_n(x_B)\ket{n}
\end{equation} 
after a total time $\tau$. This target state is the same magnon but now localized at a position $x_B\!=\!x_A+d$ with transport distance $d$. Note that $x_A,x_B$ are chosen sufficiently far from the boundaries to avoid any finite size effects. 

The efficacy of our magnon transport protocol, parameterized by $X_0(t)$, can be quantified by the infidelity
\begin{equation}\label{eq:infmagnon}
\mathcal{I}_\tau = 1 - |\bra{\psi_B}U(\tau)\ket{\psi_A}|^2. 
\end{equation} 
Here, $U(\tau)=\mathcal{T}\exp{\left[-i\int_0^\tau H(t)dt\right]}$ is the unitary operator that solves the time-dependent Schr\"{o}dinger equation with the single spin excitation Hamiltonian $H(t)$ and $\mathcal{T}$ is the Dyson time-ordering operator. This infidelity, $\mathcal{I}_\tau$, forms the objective function which we minimize with respect to $X_0(t)$ to obtain the optimal transport control pulses. We remark that this objective function must be slightly adapted for the case of a disordered system as explained in Sec.~\ref{dirty}. 

\section{Control protocols \label{proto}}
To search for the optimal magnon transport protocols that minimize the infidelity, $\mathcal{I}_\tau$, we compare and contrast several different optimization approaches and ansatzes. Specifically, we examine the performance of pulse profiles obtained with two ``hybrid'' $\partial$P control approaches, where other methods are augmented with $\partial$P, with those obtained with a standard $\partial$P time-bin optimization. For reference we will also consider a simple linear benchmark protocol
\begin{equation}\label{eq:linbench}
X_0(t) = x_A + \frac{x_B - x_A}{\tau}t,
\end{equation}
as considered in~\cite{Vbalachandran2008}. While Eq.~\eqref{eq:linbench} is comparatively easy to implement experimentally, achieving high final target state fidelities with this protocol typically requires long (adiabatic) transport times $\tau$.

\subsection{STA Protocol}
Here, we briefly recapitulate the main idea of Ref.~\cite{Kiely2021} which forms the basis for the STA protocol ansatz we will employ. We derive optimal control protocols for a particle in a harmonic trap Eq.~\eqref{eq:singlep} with the inverse engineering method based on Lewis-Riesenfeld (LR) invariants. As demonstrated in Ref.~\cite{Kiely2021}, the approximate correspondence with the single spin excitation Hamiltonian $H$ allows these protocols to be highly effective for the magnon transport in a Heisenberg chain. 

In order to apply the inverse engineering STA method we need to find a suitable dynamical invariant $I(t)$, which is a time-dependent Hermitian operator that satisfies $\frac{\partial I(t)}{\partial t} + i[H_c(t),I(t)]=0$. For the single particle in a harmonic trap with Hamiltonian 
\begin{equation}\label{eq:hoscl}
H_c(t) = \frac{k^2}{2m} + \frac{1}{2}m\omega_0^2\left[x-X_0(t)\right]^2,
\end{equation} 
a quadratic in momentum LR invariant~\cite{Lewis1969, Torrontegui2011} is given by 
\begin{equation}
I(t) = \frac{1}{2m} (k-m \dot{\alpha})^2+\frac{1}{2}m \omega_0^2 \left(x-\alpha\right)^2.
\end{equation} 
The time dependent function $\alpha\!\equiv\!\alpha(t)$ must satisfy
\begin{align}
&\ddot{\alpha}+\omega_0^2 (\alpha-X_0(t)) = 0 \label{eq:lreq2}
\end{align} 
to ensure $dI/dt =0$.

To find optimal transport protocols for the harmonic trap we require that the initial and final eigenstates of $I(t)$ match with the initial $\ket{\psi_A}$ and target states $\ket{\psi_B}$ of our control problem. This can be achieved by imposing $[I_\tau(0),H_\tau(0)]\!=\![I(\tau),H(\tau)]\!=\!0$ which results in the following additional boundary conditions 
\begin{equation}
\alpha(0) = x_A,~~~ \alpha(\tau) = x_B,~~~ \frac{d^{n}\alpha}{dt^n}\bigg\rvert_{t=0,\tau}=0, \label{eq:bclr}
\end{equation} 
for $n\!=\!1,2$. The final shortcut protocol solutions for $X_0(t)$ are then obtained by solving the invariant equations, Eq.~\eqref{eq:lreq2}, with an arbitrary function $\alpha(t)$ that satisfies these boundary conditions. In practice we can pick any function $\alpha(t)$ which fulfil the required boundary conditions, however, as part of a hybrid $\partial P$ approach, we will parameterize a specific subset of this family of solutions as outlined in the following subsection. Note that this formalism can be extended to include a time-dependent frequency.
 
\subsection{Differentiable Programming ($\partial$P)}
To optimize the transport of the wave packet in the clean system, i.e. in the absence of disorder, we minimize the infidelity with gradient based optimization algorithms. For this we obtain the gradient of the infidelity with respect to the control parameters $\frac{d\mathcal{I}_\tau}{dX_0(t)}$ with differentiable programming~\cite{Baydin2017}. This is achieved by writing the code for the infidelity $\mathcal{I}_{\tau}$ in the python library JAX~\cite{jax2018github} and exploiting its automatic differentiation feature. Due to the use of back-propagation (widely used in deep learning~\cite{Rumelhart1986}), the gradient can be obtained very efficiently, with similar computational complexity as forward evaluation of the infidelity~\cite{BAUR1983317, Griewank89onautomatic}. This allows us to use many different control parameters and also to have flexibility in the design of the cost functional, for example allowing for the addition of penalty terms and the use of neural networks \cite{Schafer2020, Coopmans2021}.

The gradient can be used in specific optimization algorithms, which we will also refer to as update schemes. The simplest is traditional gradient descent (GD) where the control parameters are updated according to $X_0^{i+1} \!=\! X_0^{i} - \mu\nabla_{X_0}\mathcal{I}_\tau^i$, for each iteration $i$ of the learning algorithm. The hyperparameter $\mu$ is known as the learning rate and must be set manually before optimization. It is common practise to use specific decay schedules for $\mu$. We empirically determine an approximation to the optimal learning rate by scanning over a range of values from $10^{-4}$ to $10^{1}$ and choose the value of $\mu$ corresponding to the lower value of infidelity. Another, more elaborate, update scheme we employ is known under the acronym ADAM~\cite{adam} and makes, amongst other things, use of an adaptive momentum of the learning rate. We find that this generally leads to faster convergence to the optimal protocols. 

We will first consider the setting where no additional information or knowledge about the problem is used to constrain the control protocols; it starts from a random or linear protocol, and we refer to this as ``$\partial$P-free''. This approach is useful for finding new unexpected protocols, for example the jump-move-jump protocol in~\cite{Coopmans2021}, since the learning is not influenced by potential human biases, see also~\cite{Tamascelli}. A clear drawback is that it can be computationally expensive to find an optimal protocol in the large unconstrained, and possibly non-convex, optimization landscape. Specifically, we discretize the control into $M$ individual time bins $X_0(t)\!\mapsto\!\left[X_0(t_1),X_0(t_2)\cdots X_0(t_M)\right]$ of time width $\Delta t\!=\!\tau/M$ and we define $t_n\!=\!n\Delta t$~\footnote{In our simulations we take $\Delta t\!=\!0.1$ which is sufficient to ensure good convergence.}. The optimization task now reduces to finding the values for each $X_0(t_n)$ that minimize $\mathcal{I}_\tau$. 

Following from the demonstrations that hybrid control approaches can be highly effective~\cite{CampbellPRL2015, Kiely2021, AdolfoPRA}, we consider the performance of protocols which combine $\partial$P and other control methods to constrain the size of the search space. In particular, we consider a ``$\partial$P-STA'' approach where we use the analytical control pulses provided by the mapping between the single excitation spin chain and the single particle as seeds for the optimization. With GD we then search over the restricted class of STA protocols which generally allows for a very fast optimization, however, as we will see, due to the restricted search space this approach may fail where others are still effective. 

For this hybrid-approach, in what follows, we define two control parameters $C_1\!=\!X_0(\tau/4)$ and $C_2\!=\!X_0(3\tau/4)$, which correspond to the positions of the trap at times $\tau/4$ and $3\tau/4$, and obtain their derivatives, $\frac{\partial \mathcal{I}}{\partial C_{1}}$ and $\frac{\partial \mathcal{I}}{\partial C_{2}}$, with $\partial$P. We remark that this can be generalized to an arbitrary number of control parameters without any significant additional computational cost for $\partial$P. $C_1$ and $C_2$ can be used as control parameters by imposing them as additional constraints when solving the polynomial for $\alpha(t)$ in Eq.~\eqref{eq:lreq2}. The advantages are that imposing the STA protocol constraints need only be done once and the resulting fixed polynomial of $C_1$ and $C_2$ can be implemented as a differentiable function for $\partial$P. We consider the minimal polynomial $\alpha(t)\!=\!\sum_{n=1}^7 b_n t^n$ that satisfies the boundary conditions Eq.~\eqref{eq:bclr} and fixes the values $C_1$ and $C_2$. This gives a whole family of STA protocols for $X_0(t)$ in which each member corresponds to a specific configuration of the expansion coefficients, $b_n$. 

The final ansatz for the optimal protocols that we use we term ``$\partial$P-Fourier'', where we aim to combine differentiable programming with optimal control based on a truncated Fourier basis and inspired by the CRAB ansatz~\cite{Muller2021, Referee3, Referee4, Referee5}. Specifically, to fulfill the boundary conditions $X_0(0)\!=\!x_A$ and $X_0(\tau)\!=\!x_B$, we define the Fourier ansatz to be the family of protocols parameterized as 
\begin{equation}\label{eq:Fansatz}
X_{0}(t) = x_A + (x_B-x_A)\frac{t}{\tau} + \sum_{n=1}^{N_{c}} A_n \sin{\omega_n t}. 
\end{equation} 
The control parameters we wish to optimize are now the Fourier coefficients $A_n$ while we fix the frequencies $\omega_n\!=\!\frac{n\pi}{\tau}$ to be the first $n \leq N_c$ harmonics. This cutoff on the maximum frequency of the protocol reflects constraints in the experimental implementation. Although often randomised frequency components are chosen to improve convergence, we remark that we do not find this necessary for our purposes.

To then find the optimal protocols in this restricted family we use $\partial$P to compute the derivatives $\frac{\partial\mathcal{I}_\tau}{\partial A_n}$ and employ them in standard gradient based optimization algorithms such as the GD algorithm defined before. The Fourier coefficients are updated from their initial values using a gradient descent update scheme with an adaptive step size. This can be done for a large number of coefficients since the gradient is calculated via differential programming. This involves the use of reverse mode automatic differentiation (back propagation) where the gradient is calculated using the chain rule in a reverse order.  Due to back-propagation, we can scale to thousands of Fourier coefficients without significantly increasing the computational cost.

In what follows we will examine the efficacy of these ansatzes where throughout: $\partial$P-free refers to the case where no constraints on the functional form of the control pulse are enforced, $\partial$P-STA corresponds to optimising over the shortcut-to-adiabaticity protocols constrained by Eqs.~\eqref{eq:lreq2} and \eqref{eq:bclr}, and finally $\partial$P-Fourier denotes the use of Eq.~\eqref{eq:Fansatz}.

\section{Optimal transport in a disorder-free chain \label{clean}}
We begin by exploring the effectiveness of these control protocols applied to transport in the ``clean'' chain, which has been extensively considered in the literature ~\cite{Wang2016,Lee2018,GrossJPB, glaser2015, Koch2022, Muller2021, GabrielePRB, Bukov, Referee1,Referee4,Referee5,Referee6, Referee7, Referee8, Referee9, Referee10, Referee11, Referee12, Referee13}. This will allow us to first test the performance in an idealized setting, providing benchmark and reference protocols for when disorder is introduced in Sec.~\ref{dirty}. Furthermore, we establish that while the STA protocols have an effective speed limit~\cite{Kiely2021}, the $\partial$P method can achieve transport velocities close to the maximum magnon group velocity of $v_{g}\!=\!2J$.

Our aim is to minimize the infidelity of the magnon transport $\mathcal{I}_\tau$ in Eq.~\eqref{eq:infmagnon} and find the optimal control protocols for the trapping center $X_0(t)$. For this task we apply $\partial$P in combination with the protocols described in the last section. To define the external constraints of the control problem we fix the total transport distance $d\!=\!50$ and choose four different total transport times $\tau\!=\!\{40,60,80,100\} J^{-1}$.  To obtain the optimal protocols we minimize with the gradient descent algorithm for a maximum of 200 update steps with an empirically determined learning rate. The minimization is stopped when an infidelity lower than $\mathcal{O}(10^{-3})$ is reached.

\begin{figure}[t]
    \centering
    \includegraphics[width=1.0\linewidth]{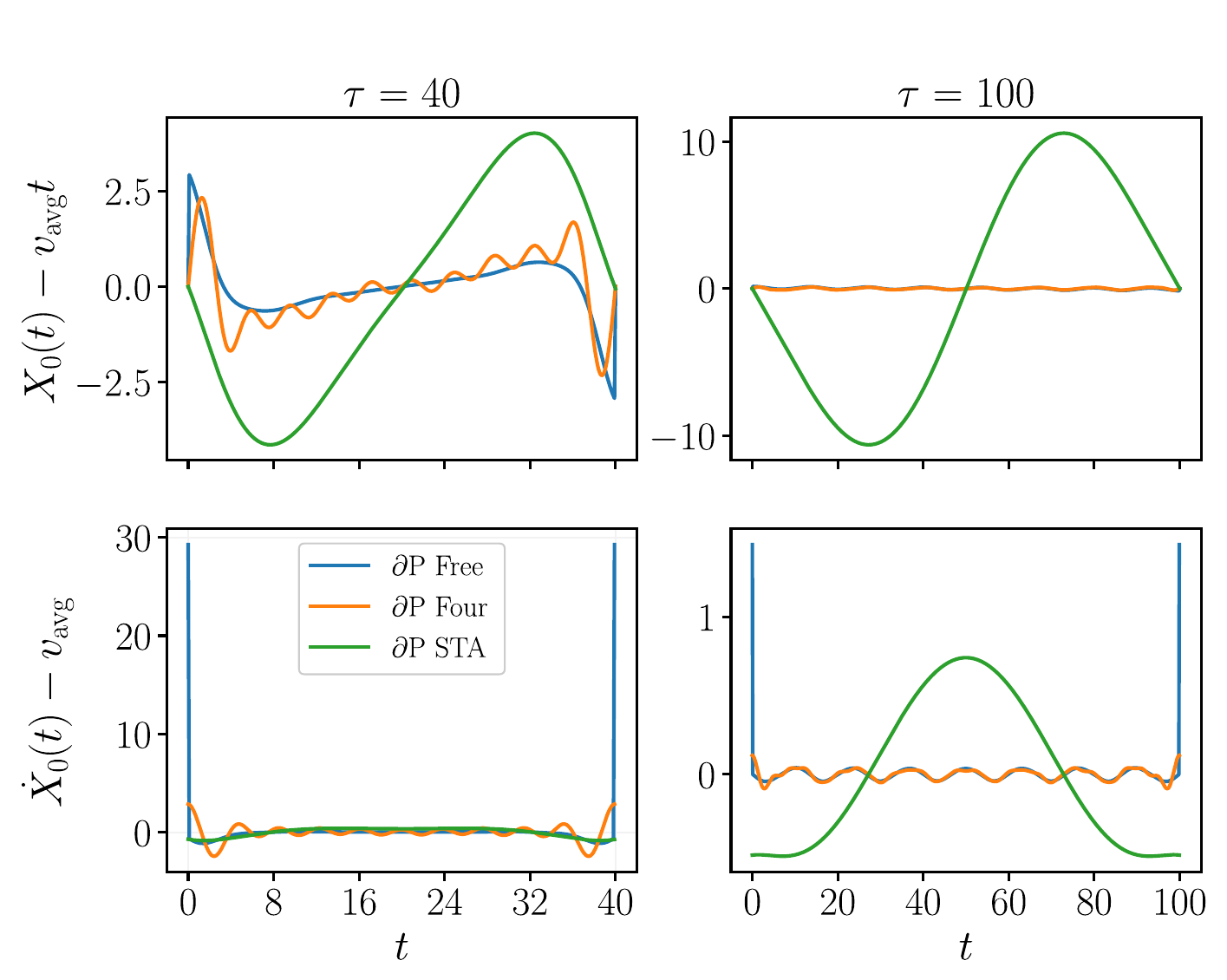}
    \caption{Disorder-free optimization results obtained with $\partial$P. The three different ansatzes are free optimization i.e. no constraints (blue), Fourier series with $N_c=\tau/2$ (orange) and the STA protocol parameterized by $C_1$ and $C_2$ (green). The results are shown relative to the linear protocol for two operation durations. Fidelities and other operations times are reported in Table. \ref{tb:res_clean}. The system parameters are chosen as $N=251$, $d=50$, $\omega_0=0.5$, and $J=1$. }
    \label{fig:cleanoptres}
\end{figure}

The results for two operation times are shown in Fig.~\ref{fig:cleanoptres} and the corresponding infidelity values are reported in Table~\ref{tb:res_clean}. Notice that above the STA speed limit, which is $\tau^*\!=\!50$ for these parameters~\cite{Kiely2021}, we are able to obtain protocols that achieve infidelities on the of order of $\mathcal{O}(10^{-3}-10^{-4})$ with all three optimization methods, while the linear reference protocol has significantly larger infidelity values (we remark the reasonably good performance of the linear ramp for $\tau\!=\!80$ is due to a special resonance effect~\cite{Kiely2021}. In the continuum case, these specific operation times of high fidelity can be expressed analytically in terms of the Fourier transform of the velocity profile of the trapping potential~\cite{Couvert2008}.). For faster protocols, $\tau\!=\!40$, the $\partial$P-Free and Fourier approaches are still able to achieve low infidelities, $\mathcal{O}(10^{-2})$, whereas $\partial$P-STA breaks down due to the fact that we have a restricted search space, although we remark it is still significantly more effective than the simple linear ramp. 

\begin{table}[t]
  \centering
  \begin{tabular}{|c|c|c|c|c|} \hline
  $\tau$ & $\partial$P-Free & $\partial$P-STA  & $\partial$P-Fourier & Linear \\\hline
  40  & 0.03832 & 0.3614   & 0.0176  &0.9845\\\hline
  60 & 0.0015   & 0.00013 & 0.0052  &0.6780\\\hline
  80 & 0.0009  &   0.00023  & 0.0003  &0.0791\\\hline
  100 & 0.0024    & 0.0002  & 0.0017  & 0.4414\\\hline
  \end{tabular}
  \caption{Infidelity $\mathcal{I}_\tau$ results for the clean system with $N=251$, $J=1$, $\omega_0$ and $d=50$. The corresponding control protocols are shown in Fig. \ref{fig:cleanoptres}.}
  \label{tb:res_clean}
\end{table}

Naturally, the shape of the optimal protocols $X_0(t)$ themselves depends strongly on the method employed, cf. Fig.~\ref{fig:cleanoptres}. The STA protocols are drastically different taking the form of a ramp up then down protocol for the velocity where they slowly accelerate to a finite velocity and then symmetrically decelerate again to zero. In contrast, the $\partial$P Free and Fourier protocols start and end with a small quench in position and in the middle oscillate with a constant average velocity. The size of the position quenches grows with decreasing transport time while the oscillations tend to become smoother with increasing time. 

\begin{figure}[t]
    \centering
    \includegraphics[width=0.45\textwidth]{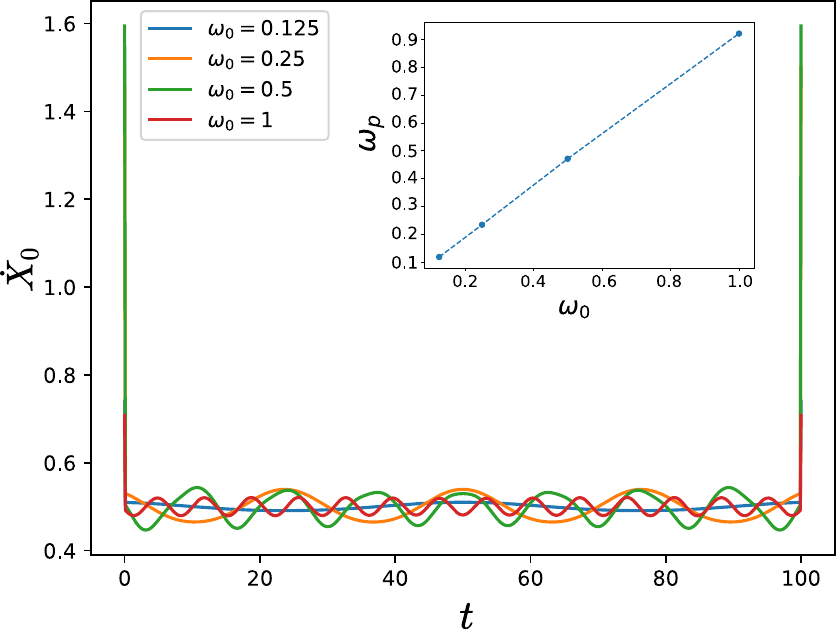}
    \caption{Main panel: velocities of the magnon transport protocols $\dot{X}_0(t)$ obtained with the $\partial$P Free approach for 4 different trapping frequencies $\omega_0$. Inset: the characteristic frequency $\omega_p$ of the protocols in the main panel versus $\omega_0$. A linear relationship $\omega_p\approx\omega_0$ is observed. Other parameters: $\tau=100$, $N=251$, $J=1$ and $d=50$. The parameters for these simulations are the same as in Fig.~\ref{fig:cleanoptres}.}
    \label{fig:analysismag}
\end{figure}

In order to analyse the frequencies, $\omega_p$, of these oscillations in Fig.~\ref{fig:analysismag}(a) we show velocity protocols $\dot{X}_0(t)$ for several trapping frequencies $\omega_0$. In the main panel we observe that the protocol frequency $\omega_p$ tends to increase with increasing $\omega_0$. From the inset we can then see that the relationship is linear with a coefficient very close to one, i.e. $\omega_p\!\approx\!\omega_0$. While such an oscillatory behavior is common in Fourier based optimal control approaches, and often these oscillations can be excluded or suppressed without affecting the effectiveness of the protocol~\cite{PatschPRX}, here we find that these oscillations are crucial to the performance of the protocol, in their absence the infidelities rise significantly. Indeed, that the unconstrained $\partial$P-Free approach also converges to protocols with these characteristic oscillations indicate their importance in the protocol. 

\subsection{Approaching the group velocity speed limit with $\partial$P}
\begin{figure}[t]
    \centering
    \includegraphics[width=1.0\linewidth]{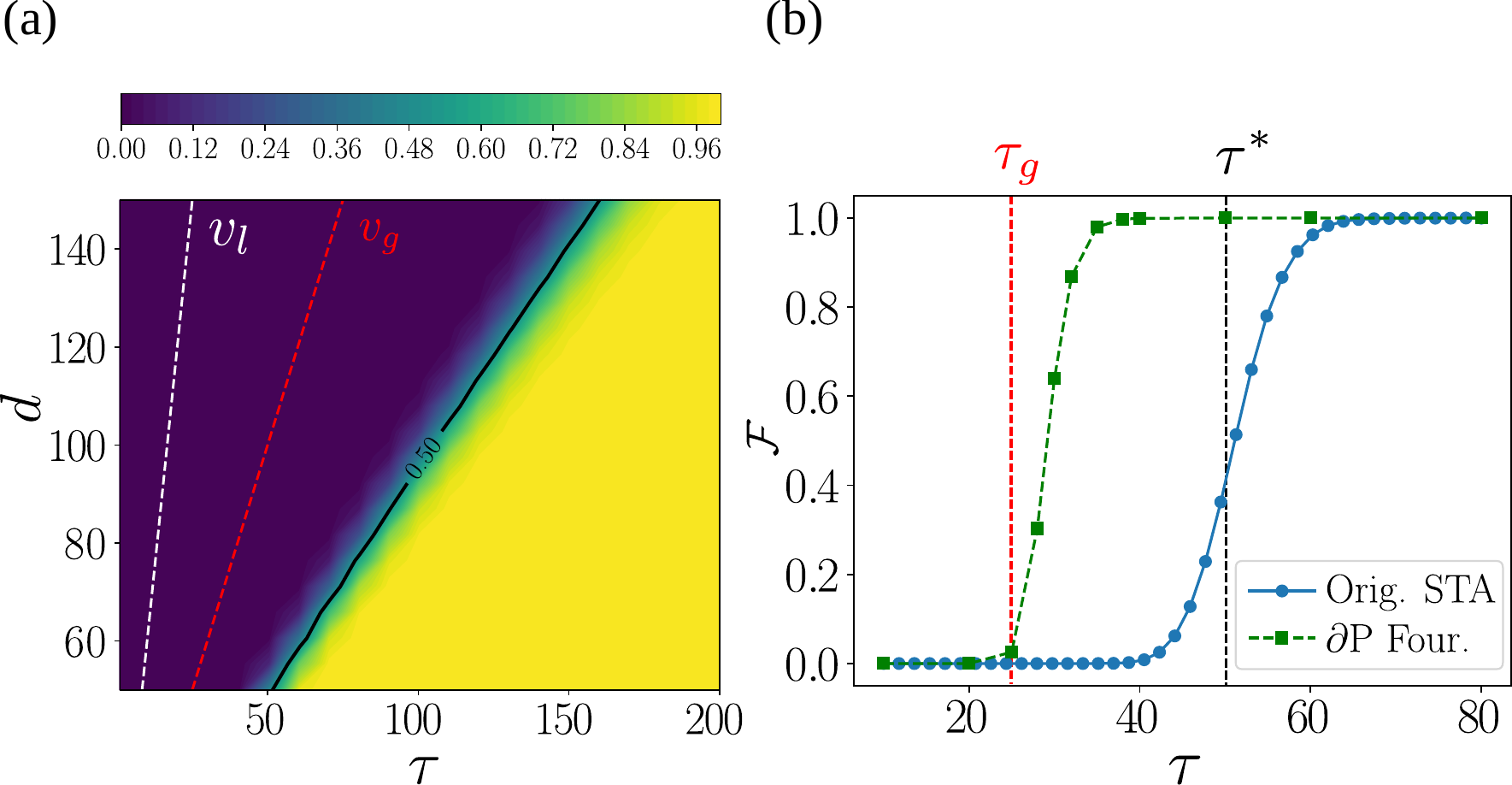}
    \caption{(a) Target state fidelity $\mathcal{F}_
    \tau$ of the standard non-parameterized STA protocol defined in Eq.~\eqref{eq:ogsta} as function of the transport distance $d$ and transport time $\tau$. The black line gives the $\mathcal{F}_\tau=0.5$ contour for which the heuristic speed limit $v_{\tau^*}=1$ is defined, the red dashed line the group velocity $v_g=2$ and the white dashed line the Lieb-Robinson velocity $v_l=6$. (b) Target state fidelity as a function of the transport time $\tau$ for the original STA protocol (blue solid line) versus the optimized $\partial$P Fourier protocols (green line). We see that the optimized $\partial$P Fourier protocols keep the fidelity at one for smaller transport times $\tau$ than the original STA protocol. The group velocity bound time is given by the red dashed line. Other parameters: $N=251$, $\omega_0=0.5$, $J=1$ and in (b) $d=50$. }
    \label{fig:heuristic_speed_limit}
\end{figure}

We can recover the speed limit for the STA protocols $\tau^*\!\approx\! d/J$ that was heuristically determined based on the original (unparameterized) STA protocol for magnon transport in~\cite{Kiely2021}. This protocol is given by
\begin{align}\label{eq:ogsta}
&X_0(t)-x_A = \nonumber \\  &d\left[6 s^5-15 s^4+10 s^3+\frac{s}{\omega_0^2 \tau^2}(1 - 3s + 2s^2)\right], \nonumber \\
\end{align}
where $s=t/\tau$. This protocol is a function of the total transport time $\tau$, the trapping frequency $\omega_0$ and the initial and target positions $x_A$ and $x_B$. In Fig.~\ref{fig:heuristic_speed_limit}(a) we show the target state fidelity $\mathcal{F}_\tau=1-\mathcal{I}_\tau$ of this protocol for a specific range of values of $d$ and $\tau$. We observe a clear lightcone-like surface with a velocity of $v_{\tau*}\!\approx\! d/J$. Below this velocity, a fidelity of at least $\mathcal{F}_\tau\!>\!0.5$ can be obtained whereas for faster protocols the fidelity drops quickly to zero. We see that this speed limit is still far away from the lightcones obtained by the group velocity $v_{g}\!=\!2J$ and even further from the Lieb-Robinson velocity~\cite{Epstein_2017} $v_l\!=\!6J$ which bounds the timescales on which quantum correlations can develop between different subsystems of a larger quantum system.

In order to demonstrate that we can improve on this speed limit with $\partial$P we focus on one specific distance slice $d=50$ and minimize $\mathcal{I}_\tau$ for a range of different total times $\tau$ near the STA speed limit time $\tau^*\!=\!50$. We use the Fourier ansatz with $N_c\!=\!50$ frequency components and, to push the limits of our optimization methodology, we run $500$ update steps with the ADAM update scheme. We stop the optimization when an infidelity of $\mathcal{O}(10^{-5})$ is reached. 

The resulting fidelity values as function of $\tau$ compared to the original STA protocol are shown in Fig.~\ref{fig:heuristic_speed_limit}(b). Compellingly, we see that we can push the maximum velocity very close to the group velocity time $\tau_g \!=\! d/v_g$. Taking $\mathcal{F}_\tau\!>\!0.5$ to measure the speed limit we find $\tau_{\partial P}\approx 29$ which corresponds to $v_{\partial P} \approx 1.72J$, which is very close to the group velocity $v_g$. Below $\tau_g$ the fidelity quickly drops to zero with no noticeable improvement arising from further optimization. This is in accordance with the group velocity being the maximum speed of the magnons~\cite{Ahmed2015} and also with the speed limit found in \cite{Murphy2010} with the numerical Krotov~\cite{krotovGlobalMethodsOptimal1993} optimization method for the same system but with a different form of spin excitation. 

\section{Magnon transport in the presence of disorder\label{dirty}}
We now consider the effects of disorder in the spin chain~\cite{Burgarth_2005,DeChiara2005,Vbalachandran2008, Ahmed2017, Kiely2021}, which could arise either from inhomogeneities in applied fields or defects in fabrication processes, and generally leads to localization, significantly hindering the system's controllability. We consider onsite disorder corresponding to an inhomogeneous magnetic trapping field $B_n(t)\!\mapsto\! B_n(t) + \epsilon_n$. Note that while the trapping field is time dependent for the control of the position of the magnon, the disorder itself is static and $\epsilon_n$ is uniformly distributed on the interval $\left[-\Delta, \Delta \right]$ with noise strength $\Delta$. 

\begin{figure}[t]
    \centering
    \includegraphics[width=1.0\linewidth]{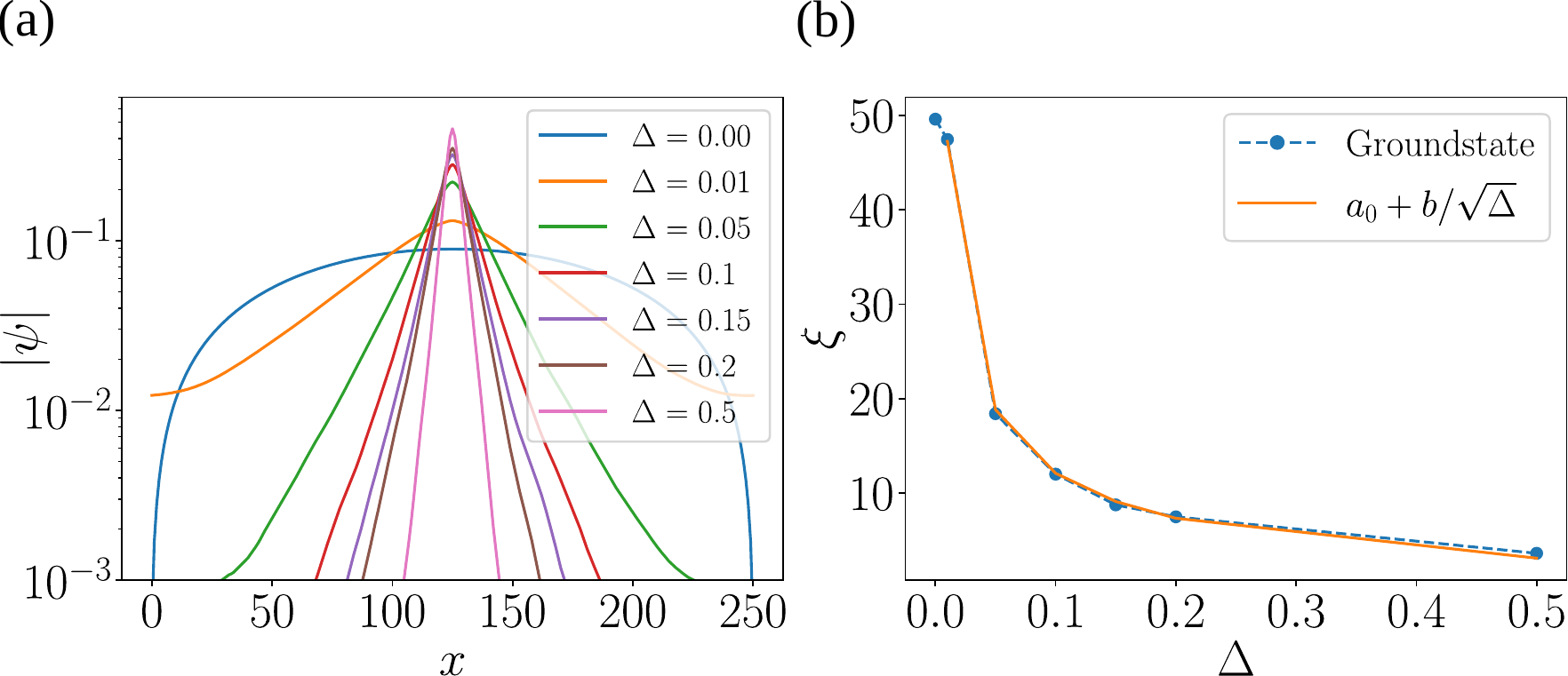}
    \caption{(a) Disorder averaged amplitude $|\psi|$ of the lowest energy eigenstate of the single spin excitation Heisenberg Hamiltonian in Eq.~\eqref{eq:seHam} with a disordered magnetic field $B_n=\epsilon_n$ which is uniformly distributed over $\left[-\Delta, \Delta \right]$. For increasing $\Delta$ the amplitude $|\psi|$ becomes localized to a smaller number of lattice sites $x$. The results are averaged over 1000 realizations and centred about  $x=125$. Other parameters: $N=251$, $\omega_0=0.5$ and $J=1$. (b) Localization length $\xi$ of the amplitudes in (a) obtained upon fitting a Laplace-like distribution $\frac{1}{\sqrt{\xi}} e^{-(x-x_0)/\xi}$. $\xi$ (blue dashed line) falls off approximately as a square root of $\Delta$ as shown by the fit (orange solid line).}
    \label{fig:localization}
\end{figure}

The main effect of the disordered magnetic field is a localization of the single spin excitation wave functions with the localization length scaling with the magnitude of the disorder, as shown in Fig.~\ref{fig:localization} for the lowest energy eigenstate of the disordered Hamiltonian $H$ for several disorder strengths. We observe that the localization length $\xi$ decreases approximately as $\xi\sim 1/\sqrt{\Delta}$. The wave function localization means that the transport infidelity is expected to increase when the magnon is moved a distance of at least a few disorder length scales $\xi$. From Fig.~\ref{fig:localization}(b) we see that already for $\Delta\!\sim\!0.05$ we have $\xi\!\sim\!40$, thus even for this small disorder the transport distance ($d=50$) is already greater than the localization length.

Based on the results from the previous section, we focus on $\partial$P in combination with the Fourier ansatz in Eq.~\eqref{eq:Fansatz} and aim to find the optimal protocol(s) in two distinct and physically relevant scenarios. In the first, we focus on one specific fixed disorder realization and determine whether effective control can still be achieved. In the second, we optimize over an ensemble of different disorder realizations, to search for an optimal control strategy that on average works the best in all disordered cases with fixed magnitude.

\subsection{Fixed disorder patterns}
We consider a particular disorder realization $\bm{\epsilon}^p=(\epsilon_1^p,...,\epsilon_N^p)$ which we label by $p$. Notice that this is simply an $N$-dimensional random vector from the uniform distribution $\left[-\Delta, \Delta\right]$. For this pattern $p$ we then aim to minimize the infidelity $\mathcal{I}_\tau^p$. This infidelity measure is the same as the standard target state infidelity $\mathcal{I}_\tau$ used before but computed with the fixed disorder pattern $p$ added to the magnetic field $B_n(t)+\epsilon_n^p$ in the Hamiltonian. For the minimization we use $\mathcal{I}_\tau^p$ as the cost functional and we compute its derivatives with respect to the control parameters with $\partial$P. These derivatives are then used in combination with ADAM to update the controls $A_n$ of the Fourier ansatz. We repeat the complete optimization process for several different disorder patterns $p$ and finally average the result $\langle \mathcal{I}_\tau \rangle_\text{s}\! = \!\frac{1}{p_{\text{max}}}\sum_p \mathcal{I}_\tau^p$, where the subscript `s' indicates that the infidelity was optimized, individually, for each disorder pattern.

\begin{figure*}[t]
    \centering
    \includegraphics[width=1.0\linewidth]{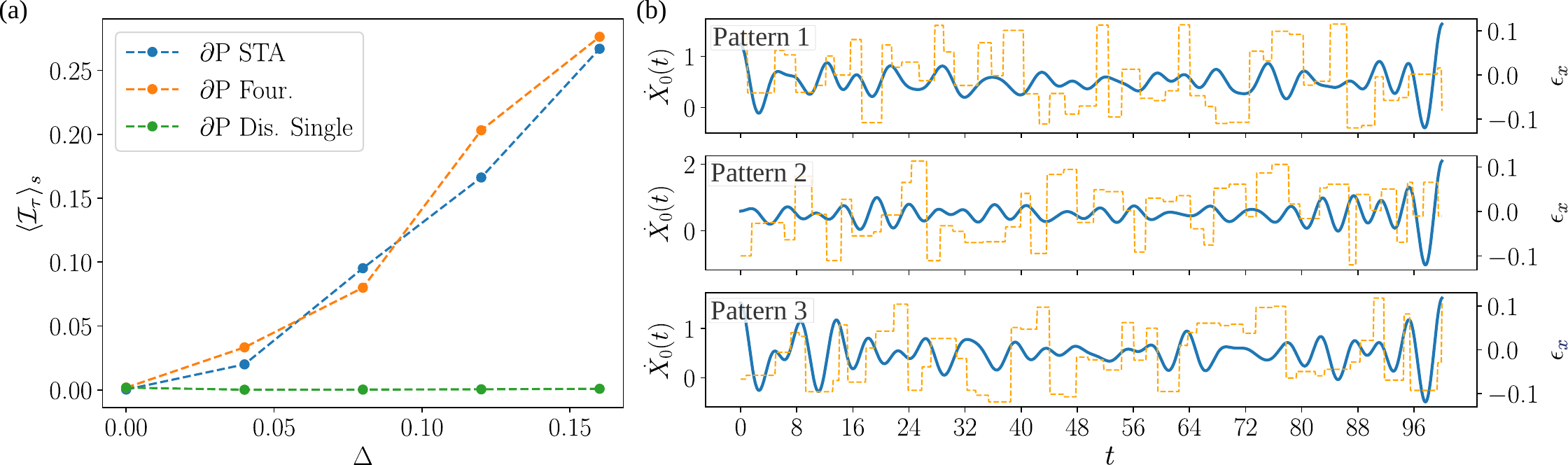}
    \caption{(a) Disorder averaged infidelity $\langle \mathcal{I}_\tau \rangle_s$ for 5 single disorder realizations versus disorder strength $\Delta$. For the $\partial$P optimization method applied to a single disorder pattern (green dashed line labeled Dis. Single), we optimized the $\mathcal{I}_\tau^p$ for each pattern $p$ individually before computing the average $\langle \mathcal{I}_\tau \rangle_s$. This optimization method is able to keep the infidelity near zero while the infidelity of the benchmark protocols obtained for optimization in the clean system increases quadratically. We used ADAM with 500 update steps on a Fourier ansatz of $N_c=50$ frequency components with external constraints $(\tau, d)=(100,50)$. All other parameters are the same as in Fig.~\ref{fig:localization}. (b) The obtained velocity protocols (left y-axis) with the $\langle \mathcal{I}_\tau \rangle_s$ method for 3 specific disorder patterns at $\Delta=0.12$. The disorder value $\epsilon_x$ where $x=\text{int}\left[X_0(t)\right]$ is plotted in orange with the scale on the right y-axis. For each different pattern we obtain a different optimal transport protocol.}
    \label{fig:single_real}
\end{figure*}
In Fig.~\ref{fig:single_real}(a) we show the results for these optimizations as a function of the disorder strength $\Delta$. We also show benchmark results of the optimal protocols obtained with the $\partial$P-STA and $\partial$P-Fourier methods in the clean system. We observe that the disorder $\partial$P optimization method (green line) is able to keep the infidelity $\langle \mathcal{I}_\tau \rangle_\text{s}$ close to zero (on the order of $10^{-3}$ and smaller). The infidelity $\langle \mathcal{I}_\tau \rangle_\text{s}$ of the benchmark protocols (blue and orange lines) for the same fixed disorder patterns is found to be increasing approximately quadratically with disorder strength. This shows that our $\partial$P optimization method is still able to obtain high fidelity control protocols in the presence of a specific disorder pattern. 

This means that disorder in the setup itself is not immediately an issue as long as a suitable control protocol, designed for the specific disorder realization, is used. Note, however, that to find these protocols the learning algorithms have used information about the exact disorder patterns, since the computed infidelity and the gradients implicitly depend on the disorder. As such, our $\partial$P ``model-based" approach only works when the exact form of the pattern is known or can be derived from experimental measurements. That said, the fact that suitable control protocols exist directly motivates the use of other gradient-free optimization approaches. We refer to the conclusion for a further discussion on this.

To try to understand how the optimization method is able to achieve these low infidelity values despite the presence of disorder, we show in Fig.~\ref{fig:single_real}(b) the optimal velocity protocols $\dot{X}_0(t)$ for three specific disorder patterns. The obtained optimal protocols $\dot{X}_0(t)$ are different for each different disorder pattern. This implies that during the optimization process the machine indeed learns about the specific disorder realization and finds a way to correct for it. This is likely due to the fact that the information of a fixed pattern $p$ is directly encoded in the derivatives $\frac{\partial \mathcal{I}_\tau^p}{\partial A_n}$ that the optimizer gets. The strategies being different, however, also means that there is likely no universal strategy that obtains a low infidelity value for any disorder realization, which we elucidate in the following subsection.

\subsection{Ensemble average disorder}
In the second scenario we consider $\partial$P optimization when we do not know the exact disorder realizations and only have access to the magnitude of the noise. For this we take the disorder averaged infidelity $\langle \mathcal{I}_\tau \rangle$ as the figure of merit to minimize with $\partial$P via batch gradient descent. We first fix a set of 200 different disorder realizations, divided up into 20 individual batches of 10 realizations. For each batch we compute the disorder averaged gradients $\frac{\partial \langle \mathcal{I}_\tau \rangle}{\partial A_n}$ and use them to update the control protocol with ADAM. We do this cyclically which means we start from the first batch do an update and then move on to the second batch. When all batches have had one update we have completed one learning `episode' and repeat the cycle. During this process $\langle \mathcal{I}_\tau \rangle$ averaged over all the 200 disorder patterns goes gradually down until convergence is reached.

\begin{figure}[t]
    \centering
    \includegraphics[width=1.00\linewidth]{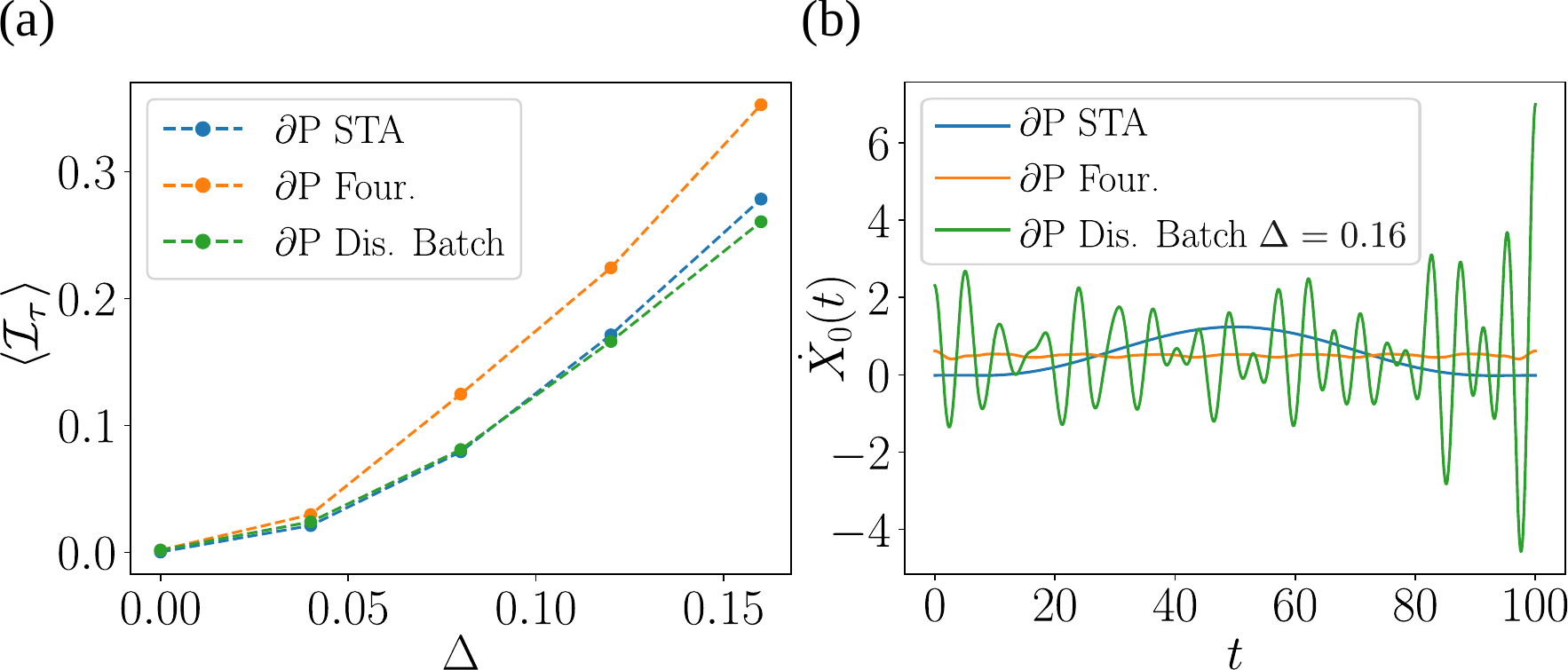}
    \caption{(a) Disorder averaged infidelity $\langle \mathcal{I}_\tau \rangle$ versus disorder strength $\Delta$ for protocols obtained with the $\partial$P batch optimization method (green dashed line) compared to the optimal clean protocols. Results are averaged over 500 disorder realizations and all the model parameters are the same as in Fig.~\ref{fig:localization}. The external constraints are set to $(\tau,d)=(100,50)$. All the protocols increase approximately quadratically with disorder strength and the batch optimization slightly outperforms the benchmark protocols. (b) The optimal velocity protocol $\dot{X}_0(t)$ (green line) obtained with the batch optimization method at $\Delta=0.16$. We have also shown the optimal protocols for the disorder free system. The batch optimization protocol has much more drastic and rapid changes in velocity compared to the optimal clean protocols.}
    \label{fig:batch_real}
\end{figure}

In Fig.~\ref{fig:batch_real}(a) we show the resulting averaged infidelity $\langle \mathcal{I}_\tau \rangle$ values for the optimal protocols obtained with batch gradient descent for 49.5 episodes.  Note that we first optimized for 24.5 episodes on one fixed set of 200 disorder realizations and then 25 episodes on a different set of 200 disorder realizations. Here 0.5 episode means that we stopped after 100 out of the 200 disorder realizations. For these $\langle \mathcal{I}_\tau \rangle$ values we have used a new set of disorder realizations which were not used during the optimization process. This gets rid of any optimization bias to particular disorder realizations and we can fairly evaluate the performance. We see that now the optimization method is not able to keep the infidelity close to zero for increasing disorder strength $\Delta$. Instead it increases quadratically similarly to the benchmark protocols, with the batch optimization method providing a negligible increase in performance compared to the other protocols. We remark this behavior is qualitatively consistent with previous studies focusing the effect of parameter fluctuations~\cite{Referee1,Referee2}.

As a final remark we note that although the batch optimization method does not give a significant improvement in infidelity compared to the STA protocols,  the form of the pulses are  different between the approaches. We show in Fig.~\ref{fig:batch_real}(b) the optimal velocity protocol obtained at a disorder strength of $\Delta\!=\!0.16$ compared to the optimal clean Fourier and STA protocols. We observe that the batch optimization protocol has much more rapid changes in the velocity and also obtains higher speeds, but nevertheless performs comparably to these protocols.

\section{Conclusion and Outlook \label{conc}}
We have examined schemes to achieve high-fidelity transport of magnons in clean and disordered Heisenberg spin chains. For this optimization task we have exploited a hybrid numerical-analytical approach employing different types of analytical ansatz for the protocols, optimized with differentiable programming ($\partial$P). In the absence of disorder, effective control can be achieved with all techniques, however, certain limiting factors relating to the maximum speed emerge, with the $\partial$P-Fourier able to approach the maximum group velocity. The relevance of external physical parameters, in particular the frequency of the trapping potential used, was shown to manifest in the optimized protocols. 

Introducing disorder, we showed that it is still possible to achieve near perfect optimal magnon transport for a fixed onsite disorder pattern, thus demonstrating that the presence of disorder is, in and of itself, not a limiting factor in using quantum systems for communication. However, we also established that even with optimization one cannot achieve a ``one size fits all'' protocol if only the disorder strength, and not the precise pattern, is known. We focussed explictly on the common assumption of Gaussian initial states. While this is useful for the approximate STA protocol, it is not critical for the implementation of the other protocols. We expect that other initial states will not qualitatively change our results. These results naturally lead to the question if the same protocols (or protocols with same performance) can be found with non-gradient based optimization approaches, such as
reinforcement learning~\cite{Sutton2018, Bukov, zhangWhenDoesReinforcement2019, Erdman2021, Erdman2022}
or natural evolution strategies~\cite{JMLR:v15:wierstra14a, Coopmans2021}. This will be particularly intriguing for the fixed disorder pattern results, as this potentially eliminates the need to know the exact disorder pattern. 

From an optimization perspective, several additional interesting avenues could be investigated. For example how the control landscape properties (smoothness, convexity, etc.)~\cite{Day_2019} for the different protocol ansatzes change with the presence of disorder. It is also relevant to explore potential relationships between the complexity of the pulse, e.g. in terms of its spectral bandwidth, and the disorder. However, to explore these kind of questions about the effects of disorder often many computationally expensive numerical simulations must be performed. To resolve this, and bring the computational complexity down, a final interesting question would be to see if one can train a neural network to predict the value of the cost function as was explored for a different problem in~\cite{dalgaard2021predicting}. 

\section*{Acknowledgements}
We thank G. Kells for helpful discussion regarding this project. L.C. acknowledges Science Foundation Ireland for financial support through Career Development Award 15/CDA/3240. G.D.C. acknowledges support by the UK EPSRC EP/S02994X/1. A.K. and S.C. acknowledge support from the Science Foundation Ireland Starting Investigator Research Grant “SpeedDemon” No. 18/SIRG/5508.

\bibliography{refs}

\begin{thebibliography}{71}%
\makeatletter
\providecommand \@ifxundefined [1]{%
 \@ifx{#1\undefined}
}%
\providecommand \@ifnum [1]{%
 \ifnum #1\expandafter \@firstoftwo
 \else \expandafter \@secondoftwo
 \fi
}%
\providecommand \@ifx [1]{%
 \ifx #1\expandafter \@firstoftwo
 \else \expandafter \@secondoftwo
 \fi
}%
\providecommand \natexlab [1]{#1}%
\providecommand \enquote  [1]{``#1''}%
\providecommand \bibnamefont  [1]{#1}%
\providecommand \bibfnamefont [1]{#1}%
\providecommand \citenamefont [1]{#1}%
\providecommand \href@noop [0]{\@secondoftwo}%
\providecommand \href [0]{\begingroup \@sanitize@url \@href}%
\providecommand \@href[1]{\@@startlink{#1}\@@href}%
\providecommand \@@href[1]{\endgroup#1\@@endlink}%
\providecommand \@sanitize@url [0]{\catcode `\\12\catcode `\$12\catcode
  `\&12\catcode `\#12\catcode `\^12\catcode `\_12\catcode `\%12\relax}%
\providecommand \@@startlink[1]{}%
\providecommand \@@endlink[0]{}%
\providecommand \url  [0]{\begingroup\@sanitize@url \@url }%
\providecommand \@url [1]{\endgroup\@href {#1}{\urlprefix }}%
\providecommand \urlprefix  [0]{URL }%
\providecommand \Eprint [0]{\href }%
\providecommand \doibase [0]{http://dx.doi.org/}%
\providecommand \selectlanguage [0]{\@gobble}%
\providecommand \bibinfo  [0]{\@secondoftwo}%
\providecommand \bibfield  [0]{\@secondoftwo}%
\providecommand \translation [1]{[#1]}%
\providecommand \BibitemOpen [0]{}%
\providecommand \bibitemStop [0]{}%
\providecommand \bibitemNoStop [0]{.\EOS\space}%
\providecommand \EOS [0]{\spacefactor3000\relax}%
\providecommand \BibitemShut  [1]{\csname bibitem#1\endcsname}%
\let\auto@bib@innerbib\@empty
\bibitem [{\citenamefont {Zoller}\ \emph {et~al.}(2005)\citenamefont {Zoller}
  \emph {et~al.}}]{Zoller2005}%
  \BibitemOpen
  \bibfield  {author} {\bibinfo {author} {\bibfnamefont {P}~\bibnamefont
  {Zoller}} \emph {et~al.},\ }\bibfield  {title} {\enquote {\bibinfo {title}
  {Quantum information processing and communication},}\ }\href {\doibase
  10.1140/epjd/e2005-00251-1} {\bibfield  {journal} {\bibinfo  {journal} {Eur.
  Phys. J. D}\ }\textbf {\bibinfo {volume} {36}},\ \bibinfo {pages} {203--228}
  (\bibinfo {year} {2005})}\BibitemShut {NoStop}%
\bibitem [{\citenamefont {Kimble}(2008)}]{Kimble2008}%
  \BibitemOpen
  \bibfield  {author} {\bibinfo {author} {\bibfnamefont {H.~J.}\ \bibnamefont
  {Kimble}},\ }\bibfield  {title} {\enquote {\bibinfo {title} {The quantum
  internet},}\ }\href {\doibase 10.1038/nature07127} {\bibfield  {journal}
  {\bibinfo  {journal} {Nature}\ }\textbf {\bibinfo {volume} {453}},\ \bibinfo
  {pages} {1023--1030} (\bibinfo {year} {2008})}\BibitemShut {NoStop}%
\bibitem [{\citenamefont {Majer}\ and\ \citenamefont
  {et~al}(2007)}]{Majer2007}%
  \BibitemOpen
  \bibfield  {author} {\bibinfo {author} {\bibfnamefont {J.}~\bibnamefont
  {Majer}}\ and\ \bibinfo {author} {\bibnamefont {et~al}},\ }\bibfield  {title}
  {\enquote {\bibinfo {title} {Coupling superconducting qubits via a cavity
  bus},}\ }\href {\doibase 10.1038/nature06184} {\bibfield  {journal} {\bibinfo
   {journal} {Nature}\ }\textbf {\bibinfo {volume} {449}},\ \bibinfo {pages}
  {443--447} (\bibinfo {year} {2007})}\BibitemShut {NoStop}%
\bibitem [{\citenamefont {\ifmmode~\mbox{\c{C}}\else \c{C}\fi{}akmak}\ \emph
  {et~al.}(2019)\citenamefont {\ifmmode~\mbox{\c{C}}\else \c{C}\fi{}akmak},
  \citenamefont {Campbell}, \citenamefont {Vacchini}, \citenamefont
  {M\"ustecapl\ifmmode \imath \else \i \fi{}o\ifmmode~\breve{g}\else
  \u{g}\fi{}lu},\ and\ \citenamefont {Paternostro}}]{BarisPRA}%
  \BibitemOpen
  \bibfield  {author} {\bibinfo {author} {\bibfnamefont {B.}~\bibnamefont
  {\ifmmode~\mbox{\c{C}}\else \c{C}\fi{}akmak}}, \bibinfo {author}
  {\bibfnamefont {Steve}\ \bibnamefont {Campbell}}, \bibinfo {author}
  {\bibfnamefont {Bassano}\ \bibnamefont {Vacchini}}, \bibinfo {author}
  {\bibfnamefont {\"Ozg\"ur~E.}\ \bibnamefont {M\"ustecapl\ifmmode \imath \else
  \i \fi{}o\ifmmode~\breve{g}\else \u{g}\fi{}lu}}, \ and\ \bibinfo {author}
  {\bibfnamefont {Mauro}\ \bibnamefont {Paternostro}},\ }\bibfield  {title}
  {\enquote {\bibinfo {title} {Robust multipartite entanglement generation via
  a collision model},}\ }\href {\doibase 10.1103/PhysRevA.99.012319} {\bibfield
   {journal} {\bibinfo  {journal} {Phys. Rev. A}\ }\textbf {\bibinfo {volume}
  {99}},\ \bibinfo {pages} {012319} (\bibinfo {year} {2019})}\BibitemShut
  {NoStop}%
\bibitem [{\citenamefont {Bose}(2003)}]{Bose2003}%
  \BibitemOpen
  \bibfield  {author} {\bibinfo {author} {\bibfnamefont {S.}~\bibnamefont
  {Bose}},\ }\bibfield  {title} {\enquote {\bibinfo {title} {Quantum
  communication through an unmodulated spin chain},}\ }\href {\doibase
  10.1103/PhysRevLett.91.207901} {\bibfield  {journal} {\bibinfo  {journal}
  {Phys. Rev. Lett.}\ }\textbf {\bibinfo {volume} {91}},\ \bibinfo {pages}
  {207901} (\bibinfo {year} {2003})}\BibitemShut {NoStop}%
\bibitem [{\citenamefont {Duan}\ \emph {et~al.}(2003)\citenamefont {Duan},
  \citenamefont {Demler},\ and\ \citenamefont {Lukin}}]{Demler2003}%
  \BibitemOpen
  \bibfield  {author} {\bibinfo {author} {\bibfnamefont {L.-M.}\ \bibnamefont
  {Duan}}, \bibinfo {author} {\bibfnamefont {E.}~\bibnamefont {Demler}}, \ and\
  \bibinfo {author} {\bibfnamefont {M.~D.}\ \bibnamefont {Lukin}},\ }\bibfield
  {title} {\enquote {\bibinfo {title} {Controlling spin exchange interactions
  of ultracold atoms in optical lattices},}\ }\href {\doibase
  10.1103/PhysRevLett.91.090402} {\bibfield  {journal} {\bibinfo  {journal}
  {Phys. Rev. Lett.}\ }\textbf {\bibinfo {volume} {91}},\ \bibinfo {pages}
  {090402} (\bibinfo {year} {2003})}\BibitemShut {NoStop}%
\bibitem [{\citenamefont {Romito}\ \emph {et~al.}(2005)\citenamefont {Romito},
  \citenamefont {Fazio},\ and\ \citenamefont {Bruder}}]{Romito2005}%
  \BibitemOpen
  \bibfield  {author} {\bibinfo {author} {\bibfnamefont {A.}~\bibnamefont
  {Romito}}, \bibinfo {author} {\bibfnamefont {R.}~\bibnamefont {Fazio}}, \
  and\ \bibinfo {author} {\bibfnamefont {C.}~\bibnamefont {Bruder}},\
  }\bibfield  {title} {\enquote {\bibinfo {title} {Solid-state quantum
  communication with josephson arrays},}\ }\href {\doibase
  10.1103/PhysRevB.71.100501} {\bibfield  {journal} {\bibinfo  {journal} {Phys.
  Rev. B}\ }\textbf {\bibinfo {volume} {71}},\ \bibinfo {pages} {100501(R)}
  (\bibinfo {year} {2005})}\BibitemShut {NoStop}%
\bibitem [{\citenamefont {Cappellaro}\ \emph {et~al.}(2007)\citenamefont
  {Cappellaro}, \citenamefont {Ramanathan},\ and\ \citenamefont
  {Cory}}]{Cappellaro2007}%
  \BibitemOpen
  \bibfield  {author} {\bibinfo {author} {\bibfnamefont {P.}~\bibnamefont
  {Cappellaro}}, \bibinfo {author} {\bibfnamefont {C.}~\bibnamefont
  {Ramanathan}}, \ and\ \bibinfo {author} {\bibfnamefont {D.~G.}\ \bibnamefont
  {Cory}},\ }\bibfield  {title} {\enquote {\bibinfo {title} {Dynamics and
  control of a quasi-one-dimensional spin system},}\ }\href {\doibase
  10.1103/PhysRevA.76.032317} {\bibfield  {journal} {\bibinfo  {journal} {Phys.
  Rev. A}\ }\textbf {\bibinfo {volume} {76}},\ \bibinfo {pages} {032317}
  (\bibinfo {year} {2007})}\BibitemShut {NoStop}%
\bibitem [{\citenamefont {Hild}\ \emph {et~al.}(2014)\citenamefont {Hild},
  \citenamefont {Fukuhara}, \citenamefont {Schau\ss{}}, \citenamefont {Zeiher},
  \citenamefont {Knap}, \citenamefont {Demler}, \citenamefont {Bloch},\ and\
  \citenamefont {Gross}}]{Hild2014}%
  \BibitemOpen
  \bibfield  {author} {\bibinfo {author} {\bibfnamefont {S.}~\bibnamefont
  {Hild}}, \bibinfo {author} {\bibfnamefont {T.}~\bibnamefont {Fukuhara}},
  \bibinfo {author} {\bibfnamefont {P.}~\bibnamefont {Schau\ss{}}}, \bibinfo
  {author} {\bibfnamefont {J.}~\bibnamefont {Zeiher}}, \bibinfo {author}
  {\bibfnamefont {M.}~\bibnamefont {Knap}}, \bibinfo {author} {\bibfnamefont
  {E.}~\bibnamefont {Demler}}, \bibinfo {author} {\bibfnamefont
  {I.}~\bibnamefont {Bloch}}, \ and\ \bibinfo {author} {\bibfnamefont
  {C.}~\bibnamefont {Gross}},\ }\bibfield  {title} {\enquote {\bibinfo {title}
  {Far-from-equilibrium spin transport in heisenberg quantum magnets},}\ }\href
  {\doibase 10.1103/PhysRevLett.113.147205} {\bibfield  {journal} {\bibinfo
  {journal} {Phys. Rev. Lett.}\ }\textbf {\bibinfo {volume} {113}},\ \bibinfo
  {pages} {147205} (\bibinfo {year} {2014})}\BibitemShut {NoStop}%
\bibitem [{\citenamefont {Qiao}\ \emph {et~al.}(2020)\citenamefont {Qiao},
  \citenamefont {Kandel}, \citenamefont {Deng}, \citenamefont {Fallahi},
  \citenamefont {Gardner}, \citenamefont {Manfra}, \citenamefont {Barnes},\
  and\ \citenamefont {Nichol}}]{Qiao2020}%
  \BibitemOpen
  \bibfield  {author} {\bibinfo {author} {\bibfnamefont {H.}~\bibnamefont
  {Qiao}}, \bibinfo {author} {\bibfnamefont {Y.~P.}\ \bibnamefont {Kandel}},
  \bibinfo {author} {\bibfnamefont {K.}~\bibnamefont {Deng}}, \bibinfo {author}
  {\bibfnamefont {S.}~\bibnamefont {Fallahi}}, \bibinfo {author} {\bibfnamefont
  {G.~C.}\ \bibnamefont {Gardner}}, \bibinfo {author} {\bibfnamefont {M.~J.}\
  \bibnamefont {Manfra}}, \bibinfo {author} {\bibfnamefont {E.}~\bibnamefont
  {Barnes}}, \ and\ \bibinfo {author} {\bibfnamefont {J.~M.}\ \bibnamefont
  {Nichol}},\ }\bibfield  {title} {\enquote {\bibinfo {title} {Coherent
  multispin exchange coupling in a quantum-dot spin chain},}\ }\href {\doibase
  10.1103/PhysRevX.10.031006} {\bibfield  {journal} {\bibinfo  {journal} {Phys.
  Rev. X}\ }\textbf {\bibinfo {volume} {10}},\ \bibinfo {pages} {031006}
  (\bibinfo {year} {2020})}\BibitemShut {NoStop}%
\bibitem [{\citenamefont {Balachandran}\ and\ \citenamefont
  {Gong}(2008)}]{Vbalachandran2008}%
  \BibitemOpen
  \bibfield  {author} {\bibinfo {author} {\bibfnamefont {V.}~\bibnamefont
  {Balachandran}}\ and\ \bibinfo {author} {\bibfnamefont {J.}~\bibnamefont
  {Gong}},\ }\bibfield  {title} {\enquote {\bibinfo {title} {Adiabatic quantum
  transport in a spin chain with a moving potential},}\ }\href {\doibase
  10.1103/PhysRevA.77.012303} {\bibfield  {journal} {\bibinfo  {journal} {Phys.
  Rev. A}\ }\textbf {\bibinfo {volume} {77}},\ \bibinfo {pages} {012303}
  (\bibinfo {year} {2008})}\BibitemShut {NoStop}%
\bibitem [{\citenamefont {Eckert}\ \emph {et~al.}(2007)\citenamefont {Eckert},
  \citenamefont {Romero-Isart},\ and\ \citenamefont {Sanpera}}]{Eckert2007}%
  \BibitemOpen
  \bibfield  {author} {\bibinfo {author} {\bibfnamefont {K.}~\bibnamefont
  {Eckert}}, \bibinfo {author} {\bibfnamefont {O.}~\bibnamefont
  {Romero-Isart}}, \ and\ \bibinfo {author} {\bibfnamefont {A.}~\bibnamefont
  {Sanpera}},\ }\bibfield  {title} {\enquote {\bibinfo {title} {Efficient
  quantum state transfer in spin chains via adiabatic passage},}\ }\href
  {\doibase 10.1088/1367-2630/9/5/155} {\bibfield  {journal} {\bibinfo
  {journal} {New J. Phys.}\ }\textbf {\bibinfo {volume} {9}},\ \bibinfo {pages}
  {155--155} (\bibinfo {year} {2007})}\BibitemShut {NoStop}%
\bibitem [{\citenamefont {Kiely}\ and\ \citenamefont
  {Campbell}(2021)}]{Kiely2021}%
  \BibitemOpen
  \bibfield  {author} {\bibinfo {author} {\bibfnamefont {A.}~\bibnamefont
  {Kiely}}\ and\ \bibinfo {author} {\bibfnamefont {S.}~\bibnamefont
  {Campbell}},\ }\bibfield  {title} {\enquote {\bibinfo {title} {Fast and
  robust magnon transport in a spin chain},}\ }\href {\doibase
  10.1088/1367-2630/abea43} {\bibfield  {journal} {\bibinfo  {journal} {New J.
  Phys.}\ }\textbf {\bibinfo {volume} {23}},\ \bibinfo {pages} {033033}
  (\bibinfo {year} {2021})}\BibitemShut {NoStop}%
\bibitem [{\citenamefont {Werschnik}\ and\ \citenamefont
  {Gross}(2007)}]{GrossJPB}%
  \BibitemOpen
  \bibfield  {author} {\bibinfo {author} {\bibfnamefont {J.}~\bibnamefont
  {Werschnik}}\ and\ \bibinfo {author} {\bibfnamefont {E.~K.~U.}\ \bibnamefont
  {Gross}},\ }\bibfield  {title} {\enquote {\bibinfo {title} {Quantum optimal
  control theory},}\ }\href {\doibase 10.1088/0953-4075/40/18/R01} {\bibfield
  {journal} {\bibinfo  {journal} {J. Phys. B}\ }\textbf {\bibinfo {volume}
  {40}},\ \bibinfo {pages} {R175} (\bibinfo {year} {2007})}\BibitemShut
  {NoStop}%
\bibitem [{\citenamefont {Glaser}\ \emph {et~al.}(2015)\citenamefont {Glaser}
  \emph {et~al.}}]{glaser2015}%
  \BibitemOpen
  \bibfield  {author} {\bibinfo {author} {\bibfnamefont {S.~J.}\ \bibnamefont
  {Glaser}} \emph {et~al.},\ }\bibfield  {title} {\enquote {\bibinfo {title}
  {Training schr{\"o}dinger's cat: quantum optimal control},}\ }\href {\doibase
  10.1140/epjd/e2015-60464-1} {\bibfield  {journal} {\bibinfo  {journal} {Eur.
  Phys. J. D}\ }\textbf {\bibinfo {volume} {69}},\ \bibinfo {pages} {279}
  (\bibinfo {year} {2015})}\BibitemShut {NoStop}%
\bibitem [{\citenamefont {Koch}\ \emph {et~al.}(2022)\citenamefont {Koch} \emph
  {et~al.}}]{Koch2022}%
  \BibitemOpen
  \bibfield  {author} {\bibinfo {author} {\bibfnamefont {C.~P.}\ \bibnamefont
  {Koch}} \emph {et~al.},\ }\bibfield  {title} {\enquote {\bibinfo {title}
  {Quantum optimal control in quantum technologies. strategic report on current
  status, visions and goals for research in europe},}\ }\href {\doibase
  10.1140/epjqt/s40507-022-00138-x} {\bibfield  {journal} {\bibinfo  {journal}
  {{EPJ} Quantum Technology}\ }\textbf {\bibinfo {volume} {9}},\ \bibinfo
  {pages} {19} (\bibinfo {year} {2022})}\BibitemShut {NoStop}%
\bibitem [{\citenamefont {M{\"u}ller}\ \emph {et~al.}(2022)\citenamefont
  {M{\"u}ller}, \citenamefont {Said}, \citenamefont {Jelezko}, \citenamefont
  {Calarco},\ and\ \citenamefont {Montangero}}]{Muller2021}%
  \BibitemOpen
  \bibfield  {author} {\bibinfo {author} {\bibfnamefont {M.~M.}\ \bibnamefont
  {M{\"u}ller}}, \bibinfo {author} {\bibfnamefont {R.~S.}\ \bibnamefont
  {Said}}, \bibinfo {author} {\bibfnamefont {F.}~\bibnamefont {Jelezko}},
  \bibinfo {author} {\bibfnamefont {T.}~\bibnamefont {Calarco}}, \ and\
  \bibinfo {author} {\bibfnamefont {S.}~\bibnamefont {Montangero}},\ }\bibfield
   {title} {\enquote {\bibinfo {title} {One decade of quantum optimal control
  in the chopped random basis},}\ }\href
  {http://iopscience.iop.org/article/10.1088/1361-6633/ac723c} {\bibfield
  {journal} {\bibinfo  {journal} {Rep. Prog. Phys.}\ }\textbf {\bibinfo
  {volume} {85}},\ \bibinfo {pages} {076001} (\bibinfo {year}
  {2022})}\BibitemShut {NoStop}%
\bibitem [{\citenamefont {Power}\ and\ \citenamefont
  {De~Chiara}(2013)}]{GabrielePRB}%
  \BibitemOpen
  \bibfield  {author} {\bibinfo {author} {\bibfnamefont {M.~J.~M.}\
  \bibnamefont {Power}}\ and\ \bibinfo {author} {\bibfnamefont
  {G.}~\bibnamefont {De~Chiara}},\ }\bibfield  {title} {\enquote {\bibinfo
  {title} {Dynamical symmetry breaking with optimal control: Reducing the
  number of pieces},}\ }\href {\doibase 10.1103/PhysRevB.88.214106} {\bibfield
  {journal} {\bibinfo  {journal} {Phys. Rev. B}\ }\textbf {\bibinfo {volume}
  {88}},\ \bibinfo {pages} {214106} (\bibinfo {year} {2013})}\BibitemShut
  {NoStop}%
\bibitem [{\citenamefont {Bukov}\ \emph {et~al.}(2018)\citenamefont {Bukov},
  \citenamefont {Day}, \citenamefont {Sels}, \citenamefont {Weinberg},
  \citenamefont {Polkovnikov},\ and\ \citenamefont {Mehta}}]{Bukov}%
  \BibitemOpen
  \bibfield  {author} {\bibinfo {author} {\bibfnamefont {M.}~\bibnamefont
  {Bukov}}, \bibinfo {author} {\bibfnamefont {A.~G.~R.}\ \bibnamefont {Day}},
  \bibinfo {author} {\bibfnamefont {D.}~\bibnamefont {Sels}}, \bibinfo {author}
  {\bibfnamefont {P.}~\bibnamefont {Weinberg}}, \bibinfo {author}
  {\bibfnamefont {A.}~\bibnamefont {Polkovnikov}}, \ and\ \bibinfo {author}
  {\bibfnamefont {P.}~\bibnamefont {Mehta}},\ }\bibfield  {title} {\enquote
  {\bibinfo {title} {Reinforcement learning in different phases of quantum
  control},}\ }\href {\doibase 10.1103/PhysRevX.8.031086} {\bibfield  {journal}
  {\bibinfo  {journal} {Phys. Rev. X}\ }\textbf {\bibinfo {volume} {8}},\
  \bibinfo {pages} {031086} (\bibinfo {year} {2018})}\BibitemShut {NoStop}%
\bibitem [{\citenamefont {Caneva}\ \emph {et~al.}(2009)\citenamefont {Caneva},
  \citenamefont {Murphy}, \citenamefont {Calarco}, \citenamefont {Fazio},
  \citenamefont {Montangero}, \citenamefont {Giovannetti},\ and\ \citenamefont
  {Santoro}}]{Referee6}%
  \BibitemOpen
  \bibfield  {author} {\bibinfo {author} {\bibfnamefont {T.}~\bibnamefont
  {Caneva}}, \bibinfo {author} {\bibfnamefont {M.}~\bibnamefont {Murphy}},
  \bibinfo {author} {\bibfnamefont {T.}~\bibnamefont {Calarco}}, \bibinfo
  {author} {\bibfnamefont {R.}~\bibnamefont {Fazio}}, \bibinfo {author}
  {\bibfnamefont {S.}~\bibnamefont {Montangero}}, \bibinfo {author}
  {\bibfnamefont {V.}~\bibnamefont {Giovannetti}}, \ and\ \bibinfo {author}
  {\bibfnamefont {G.~E.}\ \bibnamefont {Santoro}},\ }\bibfield  {title}
  {\enquote {\bibinfo {title} {Optimal control at the quantum speed limit},}\
  }\href {\doibase 10.1103/PhysRevLett.103.240501} {\bibfield  {journal}
  {\bibinfo  {journal} {Phys. Rev. Lett.}\ }\textbf {\bibinfo {volume} {103}},\
  \bibinfo {pages} {240501} (\bibinfo {year} {2009})}\BibitemShut {NoStop}%
\bibitem [{\citenamefont {Wang}\ \emph {et~al.}(2010)\citenamefont {Wang},
  \citenamefont {Bayat}, \citenamefont {Schirmer},\ and\ \citenamefont
  {Bose}}]{Referee7}%
  \BibitemOpen
  \bibfield  {author} {\bibinfo {author} {\bibfnamefont {Xiaoting}\
  \bibnamefont {Wang}}, \bibinfo {author} {\bibfnamefont {Abolfazl}\
  \bibnamefont {Bayat}}, \bibinfo {author} {\bibfnamefont {S.~G.}\ \bibnamefont
  {Schirmer}}, \ and\ \bibinfo {author} {\bibfnamefont {Sougato}\ \bibnamefont
  {Bose}},\ }\bibfield  {title} {\enquote {\bibinfo {title} {Robust
  entanglement in antiferromagnetic heisenberg chains by single-spin optimal
  control},}\ }\href {\doibase 10.1103/PhysRevA.81.032312} {\bibfield
  {journal} {\bibinfo  {journal} {Phys. Rev. A}\ }\textbf {\bibinfo {volume}
  {81}},\ \bibinfo {pages} {032312} (\bibinfo {year} {2010})}\BibitemShut
  {NoStop}%
\bibitem [{\citenamefont {Caneva}\ \emph {et~al.}(2011)\citenamefont {Caneva},
  \citenamefont {Calarco},\ and\ \citenamefont {Montangero}}]{Referee8}%
  \BibitemOpen
  \bibfield  {author} {\bibinfo {author} {\bibfnamefont {Tommaso}\ \bibnamefont
  {Caneva}}, \bibinfo {author} {\bibfnamefont {Tommaso}\ \bibnamefont
  {Calarco}}, \ and\ \bibinfo {author} {\bibfnamefont {Simone}\ \bibnamefont
  {Montangero}},\ }\bibfield  {title} {\enquote {\bibinfo {title} {Chopped
  random-basis quantum optimization},}\ }\href {\doibase
  10.1103/PhysRevA.84.022326} {\bibfield  {journal} {\bibinfo  {journal} {Phys.
  Rev. A}\ }\textbf {\bibinfo {volume} {84}},\ \bibinfo {pages} {022326}
  (\bibinfo {year} {2011})}\BibitemShut {NoStop}%
\bibitem [{\citenamefont {Poggi}\ and\ \citenamefont
  {Wisniacki}(2016)}]{Referee9}%
  \BibitemOpen
  \bibfield  {author} {\bibinfo {author} {\bibfnamefont {P.~M.}\ \bibnamefont
  {Poggi}}\ and\ \bibinfo {author} {\bibfnamefont {D.~A.}\ \bibnamefont
  {Wisniacki}},\ }\bibfield  {title} {\enquote {\bibinfo {title} {Optimal
  control of many-body quantum dynamics: Chaos and complexity},}\ }\href
  {\doibase 10.1103/PhysRevA.94.033406} {\bibfield  {journal} {\bibinfo
  {journal} {Phys. Rev. A}\ }\textbf {\bibinfo {volume} {94}},\ \bibinfo
  {pages} {033406} (\bibinfo {year} {2016})}\BibitemShut {NoStop}%
\bibitem [{\citenamefont {Gurman}\ \emph {et~al.}(2016)\citenamefont {Gurman},
  \citenamefont {Guseva},\ and\ \citenamefont {Fesko}}]{Referee10}%
  \BibitemOpen
  \bibfield  {author} {\bibinfo {author} {\bibfnamefont {Vladimir~I.}\
  \bibnamefont {Gurman}}, \bibinfo {author} {\bibfnamefont {Irina~S.}\
  \bibnamefont {Guseva}}, \ and\ \bibinfo {author} {\bibfnamefont {Oles~V.}\
  \bibnamefont {Fesko}},\ }\bibfield  {title} {\enquote {\bibinfo {title}
  {Optimization of excitation transfer in a spin chain},}\ }in\ \href {\doibase
  10.1063/1.4952193} {\emph {\bibinfo {booktitle} {{AIP} Conference
  Proceedings}}}\ (\bibinfo  {publisher} {Author(s)},\ \bibinfo {year}
  {2016})\BibitemShut {NoStop}%
\bibitem [{\citenamefont {Zhang}\ \emph {et~al.}(2016)\citenamefont {Zhang},
  \citenamefont {Shao}, \citenamefont {Hu}, \citenamefont {Zou},\ and\
  \citenamefont {Wu}}]{Referee11}%
  \BibitemOpen
  \bibfield  {author} {\bibinfo {author} {\bibfnamefont {Xiong-Peng}\
  \bibnamefont {Zhang}}, \bibinfo {author} {\bibfnamefont {Bin}\ \bibnamefont
  {Shao}}, \bibinfo {author} {\bibfnamefont {Shuai}\ \bibnamefont {Hu}},
  \bibinfo {author} {\bibfnamefont {Jian}\ \bibnamefont {Zou}}, \ and\ \bibinfo
  {author} {\bibfnamefont {Lian-Ao}\ \bibnamefont {Wu}},\ }\bibfield  {title}
  {\enquote {\bibinfo {title} {Optimal control of fast and high-fidelity
  quantum state transfer in spin-1/2 chains},}\ }\href {\doibase
  10.1016/j.aop.2016.10.020} {\bibfield  {journal} {\bibinfo  {journal} {Annals
  of Physics}\ }\textbf {\bibinfo {volume} {375}},\ \bibinfo {pages} {435--443}
  (\bibinfo {year} {2016})}\BibitemShut {NoStop}%
\bibitem [{\citenamefont {Coden}\ \emph {et~al.}(2021)\citenamefont {Coden},
  \citenamefont {G{\'{o}}mez}, \citenamefont {Ferr{\'{o}}n},\ and\
  \citenamefont {Osenda}}]{Referee12}%
  \BibitemOpen
  \bibfield  {author} {\bibinfo {author} {\bibfnamefont {D.S.~Acosta}\
  \bibnamefont {Coden}}, \bibinfo {author} {\bibfnamefont {S.S.}\ \bibnamefont
  {G{\'{o}}mez}}, \bibinfo {author} {\bibfnamefont {A.}~\bibnamefont
  {Ferr{\'{o}}n}}, \ and\ \bibinfo {author} {\bibfnamefont {O.}~\bibnamefont
  {Osenda}},\ }\bibfield  {title} {\enquote {\bibinfo {title} {Controlled
  quantum state transfer in {XX} spin chains at the quantum speed limit},}\
  }\href {\doibase 10.1016/j.physleta.2020.127009} {\bibfield  {journal}
  {\bibinfo  {journal} {Physics Letters A}\ }\textbf {\bibinfo {volume}
  {387}},\ \bibinfo {pages} {127009} (\bibinfo {year} {2021})}\BibitemShut
  {NoStop}%
\bibitem [{\citenamefont {Iversen}\ \emph {et~al.}(2020)\citenamefont
  {Iversen}, \citenamefont {Barfknecht}, \citenamefont {Foerster},\ and\
  \citenamefont {Zinner}}]{Referee13}%
  \BibitemOpen
  \bibfield  {author} {\bibinfo {author} {\bibfnamefont {M}~\bibnamefont
  {Iversen}}, \bibinfo {author} {\bibfnamefont {R~E}\ \bibnamefont
  {Barfknecht}}, \bibinfo {author} {\bibfnamefont {A}~\bibnamefont {Foerster}},
  \ and\ \bibinfo {author} {\bibfnamefont {N~T}\ \bibnamefont {Zinner}},\
  }\bibfield  {title} {\enquote {\bibinfo {title} {State transfer in an
  inhomogeneous spin chain},}\ }\href {\doibase 10.1088/1361-6455/ab9076}
  {\bibfield  {journal} {\bibinfo  {journal} {Journal of Physics B: Atomic,
  Molecular and Optical Physics}\ }\textbf {\bibinfo {volume} {53}},\ \bibinfo
  {pages} {155301} (\bibinfo {year} {2020})}\BibitemShut {NoStop}%
\bibitem [{\citenamefont {Wengert}(1964)}]{Wengert1964}%
  \BibitemOpen
  \bibfield  {author} {\bibinfo {author} {\bibfnamefont {R.~E.}\ \bibnamefont
  {Wengert}},\ }\bibfield  {title} {\enquote {\bibinfo {title} {A simple
  automatic derivative evaluation program},}\ }\href {\doibase
  10.1145/355586.364791} {\bibfield  {journal} {\bibinfo  {journal} {Commun.
  ACM}\ }\textbf {\bibinfo {volume} {7}},\ \bibinfo {pages} {463–464}
  (\bibinfo {year} {1964})}\BibitemShut {NoStop}%
\bibitem [{\citenamefont {Liao}\ \emph {et~al.}(2019)\citenamefont {Liao},
  \citenamefont {Liu}, \citenamefont {Wang},\ and\ \citenamefont
  {Xiang}}]{Liao2019}%
  \BibitemOpen
  \bibfield  {author} {\bibinfo {author} {\bibfnamefont {H.-J.}\ \bibnamefont
  {Liao}}, \bibinfo {author} {\bibfnamefont {J.-G.}\ \bibnamefont {Liu}},
  \bibinfo {author} {\bibfnamefont {L.}~\bibnamefont {Wang}}, \ and\ \bibinfo
  {author} {\bibfnamefont {T.}~\bibnamefont {Xiang}},\ }\bibfield  {title}
  {\enquote {\bibinfo {title} {Differentiable programming tensor networks},}\
  }\href {\doibase 10.1103/PhysRevX.9.031041} {\bibfield  {journal} {\bibinfo
  {journal} {Phys. Rev. X}\ }\textbf {\bibinfo {volume} {9}},\ \bibinfo {pages}
  {031041} (\bibinfo {year} {2019})}\BibitemShut {NoStop}%
\bibitem [{\citenamefont {Baydin}\ \emph {et~al.}(2018)\citenamefont {Baydin},
  \citenamefont {Pearlmutter}, \citenamefont {Radul},\ and\ \citenamefont
  {Siskind}}]{Baydin2017}%
  \BibitemOpen
  \bibfield  {author} {\bibinfo {author} {\bibfnamefont
  {At\i{}l\i{}m~G\"{u}nes}\ \bibnamefont {Baydin}}, \bibinfo {author}
  {\bibfnamefont {Barak~A.}\ \bibnamefont {Pearlmutter}}, \bibinfo {author}
  {\bibfnamefont {Alexey~Andreyevich}\ \bibnamefont {Radul}}, \ and\ \bibinfo
  {author} {\bibfnamefont {Jeffrey~Mark}\ \bibnamefont {Siskind}},\ }\bibfield
  {title} {\enquote {\bibinfo {title} {Automatic differentiation in machine
  learning: A survey},}\ }\href {http://jmlr.org/papers/v18/17-468.html}
  {\bibfield  {journal} {\bibinfo  {journal} {J. Mach. Learn. Res.}\ }\textbf
  {\bibinfo {volume} {18}},\ \bibinfo {pages} {1} (\bibinfo {year}
  {2018})}\BibitemShut {NoStop}%
\bibitem [{\citenamefont {Coopmans}\ \emph {et~al.}(2021)\citenamefont
  {Coopmans}, \citenamefont {Luo}, \citenamefont {Kells}, \citenamefont
  {Clark},\ and\ \citenamefont {Carrasquilla}}]{Coopmans2021}%
  \BibitemOpen
  \bibfield  {author} {\bibinfo {author} {\bibfnamefont {L.}~\bibnamefont
  {Coopmans}}, \bibinfo {author} {\bibfnamefont {D.}~\bibnamefont {Luo}},
  \bibinfo {author} {\bibfnamefont {G.}~\bibnamefont {Kells}}, \bibinfo
  {author} {\bibfnamefont {B.~K.}\ \bibnamefont {Clark}}, \ and\ \bibinfo
  {author} {\bibfnamefont {J.}~\bibnamefont {Carrasquilla}},\ }\bibfield
  {title} {\enquote {\bibinfo {title} {Protocol discovery for the quantum
  control of majoranas by differentiable programming and natural evolution
  strategies},}\ }\href {\doibase 10.1103/PRXQuantum.2.020332} {\bibfield
  {journal} {\bibinfo  {journal} {PRX Quantum}\ }\textbf {\bibinfo {volume}
  {2}},\ \bibinfo {pages} {020332} (\bibinfo {year} {2021})}\BibitemShut
  {NoStop}%
\bibitem [{\citenamefont {Khait}\ \emph {et~al.}(2022)\citenamefont {Khait},
  \citenamefont {Carrasquilla},\ and\ \citenamefont
  {Segal}}]{khait2021optimal}%
  \BibitemOpen
  \bibfield  {author} {\bibinfo {author} {\bibfnamefont {Ilia}\ \bibnamefont
  {Khait}}, \bibinfo {author} {\bibfnamefont {Juan}\ \bibnamefont
  {Carrasquilla}}, \ and\ \bibinfo {author} {\bibfnamefont {Dvira}\
  \bibnamefont {Segal}},\ }\bibfield  {title} {\enquote {\bibinfo {title}
  {Optimal control of quantum thermal machines using machine learning},}\
  }\href {\doibase 10.1103/PhysRevResearch.4.L012029} {\bibfield  {journal}
  {\bibinfo  {journal} {Phys. Rev. Research}\ }\textbf {\bibinfo {volume}
  {4}},\ \bibinfo {pages} {L012029} (\bibinfo {year} {2022})}\BibitemShut
  {NoStop}%
\bibitem [{\citenamefont {Rigo}\ and\ \citenamefont
  {Mitchell}(2022)}]{JonasPRR}%
  \BibitemOpen
  \bibfield  {author} {\bibinfo {author} {\bibfnamefont {Jonas~B.}\
  \bibnamefont {Rigo}}\ and\ \bibinfo {author} {\bibfnamefont {Andrew~K.}\
  \bibnamefont {Mitchell}},\ }\bibfield  {title} {\enquote {\bibinfo {title}
  {Automatic differentiable numerical renormalization group},}\ }\href
  {\doibase 10.1103/PhysRevResearch.4.013227} {\bibfield  {journal} {\bibinfo
  {journal} {Phys. Rev. Research}\ }\textbf {\bibinfo {volume} {4}},\ \bibinfo
  {pages} {013227} (\bibinfo {year} {2022})}\BibitemShut {NoStop}%
\bibitem [{\citenamefont {Schäfer}\ \emph {et~al.}(2020)\citenamefont
  {Schäfer}, \citenamefont {Kloc}, \citenamefont {Bruder},\ and\ \citenamefont
  {Lörch}}]{Schafer2020}%
  \BibitemOpen
  \bibfield  {author} {\bibinfo {author} {\bibfnamefont {F.}~\bibnamefont
  {Schäfer}}, \bibinfo {author} {\bibfnamefont {M.}~\bibnamefont {Kloc}},
  \bibinfo {author} {\bibfnamefont {C.}~\bibnamefont {Bruder}}, \ and\ \bibinfo
  {author} {\bibfnamefont {N.}~\bibnamefont {Lörch}},\ }\bibfield  {title}
  {\enquote {\bibinfo {title} {A differentiable programming method for quantum
  control},}\ }\href {\doibase 10.1088/2632-2153/ab9802} {\bibfield  {journal}
  {\bibinfo  {journal} {Mach. Learn.: Sci. Technol}\ }\textbf {\bibinfo
  {volume} {3}},\ \bibinfo {pages} {035009} (\bibinfo {year}
  {2020})}\BibitemShut {NoStop}%
\bibitem [{\citenamefont {Leung}\ \emph {et~al.}(2017)\citenamefont {Leung},
  \citenamefont {Abdelhafez}, \citenamefont {Koch},\ and\ \citenamefont
  {Schuster}}]{Leung2017}%
  \BibitemOpen
  \bibfield  {author} {\bibinfo {author} {\bibfnamefont {Nelson}\ \bibnamefont
  {Leung}}, \bibinfo {author} {\bibfnamefont {Mohamed}\ \bibnamefont
  {Abdelhafez}}, \bibinfo {author} {\bibfnamefont {Jens}\ \bibnamefont {Koch}},
  \ and\ \bibinfo {author} {\bibfnamefont {David}\ \bibnamefont {Schuster}},\
  }\bibfield  {title} {\enquote {\bibinfo {title} {Speedup for quantum optimal
  control from automatic differentiation based on graphics processing units},}\
  }\href {\doibase 10.1103/PhysRevA.95.042318} {\bibfield  {journal} {\bibinfo
  {journal} {Phys. Rev. A}\ }\textbf {\bibinfo {volume} {95}},\ \bibinfo
  {pages} {042318} (\bibinfo {year} {2017})}\BibitemShut {NoStop}%
\bibitem [{\citenamefont {Anderson}(1958)}]{Anderson1958}%
  \BibitemOpen
  \bibfield  {author} {\bibinfo {author} {\bibfnamefont {P.~W.}\ \bibnamefont
  {Anderson}},\ }\bibfield  {title} {\enquote {\bibinfo {title} {Absence of
  diffusion in certain random lattices},}\ }\href {\doibase
  10.1103/PhysRev.109.1492} {\bibfield  {journal} {\bibinfo  {journal} {Phys.
  Rev.}\ }\textbf {\bibinfo {volume} {109}},\ \bibinfo {pages} {1492--1505}
  (\bibinfo {year} {1958})}\BibitemShut {NoStop}%
\bibitem [{\citenamefont {Ahmed}(2017)}]{Ahmed2017}%
  \BibitemOpen
  \bibfield  {author} {\bibinfo {author} {\bibfnamefont {M.}~\bibnamefont
  {Ahmed}},\ }\bibfield  {title} {\enquote {\bibinfo {title} {Confined magnon
  transport in low dimensional ferromagnetic structures},}\ }\href@noop {}
  {\bibfield  {journal} {\bibinfo  {journal} {Ph.D Thesis, RMIT University,
  Melbourne, Australia}\ } (\bibinfo {year} {2017})}\BibitemShut {NoStop}%
\bibitem [{\citenamefont {Ahmed}\ and\ \citenamefont
  {Greentree}(2015)}]{Ahmed2015}%
  \BibitemOpen
  \bibfield  {author} {\bibinfo {author} {\bibfnamefont {M.~H.}\ \bibnamefont
  {Ahmed}}\ and\ \bibinfo {author} {\bibfnamefont {A.~D.}\ \bibnamefont
  {Greentree}},\ }\bibfield  {title} {\enquote {\bibinfo {title} {Guided magnon
  transport in spin chains: Transport speed and correcting for disorder},}\
  }\href {\doibase 10.1103/PhysRevA.91.022306} {\bibfield  {journal} {\bibinfo
  {journal} {Phys. Rev. A}\ }\textbf {\bibinfo {volume} {91}},\ \bibinfo
  {pages} {022306} (\bibinfo {year} {2015})}\BibitemShut {NoStop}%
\bibitem [{\citenamefont {Makin}\ \emph {et~al.}(2012)\citenamefont {Makin},
  \citenamefont {Cole}, \citenamefont {Hill},\ and\ \citenamefont
  {Greentree}}]{Makin2012}%
  \BibitemOpen
  \bibfield  {author} {\bibinfo {author} {\bibfnamefont {M.~I.}\ \bibnamefont
  {Makin}}, \bibinfo {author} {\bibfnamefont {J.~H.}\ \bibnamefont {Cole}},
  \bibinfo {author} {\bibfnamefont {C.~D.}\ \bibnamefont {Hill}}, \ and\
  \bibinfo {author} {\bibfnamefont {A.~D.}\ \bibnamefont {Greentree}},\
  }\bibfield  {title} {\enquote {\bibinfo {title} {Spin guides and spin
  splitters: Waveguide analogies in one-dimensional spin chains},}\ }\href
  {\doibase 10.1103/PhysRevLett.108.017207} {\bibfield  {journal} {\bibinfo
  {journal} {Phys. Rev. Lett.}\ }\textbf {\bibinfo {volume} {108}},\ \bibinfo
  {pages} {017207} (\bibinfo {year} {2012})}\BibitemShut {NoStop}%
\bibitem [{\citenamefont {Torrontegui}\ \emph {et~al.}(2011)\citenamefont
  {Torrontegui}, \citenamefont {Ib\'a\~nez}, \citenamefont {Chen},
  \citenamefont {Ruschhaupt}, \citenamefont {Gu\'ery-Odelin},\ and\
  \citenamefont {Muga}}]{Torrontegui2011}%
  \BibitemOpen
  \bibfield  {author} {\bibinfo {author} {\bibfnamefont {E.}~\bibnamefont
  {Torrontegui}}, \bibinfo {author} {\bibfnamefont {S.}~\bibnamefont
  {Ib\'a\~nez}}, \bibinfo {author} {\bibfnamefont {Xi}~\bibnamefont {Chen}},
  \bibinfo {author} {\bibfnamefont {A.}~\bibnamefont {Ruschhaupt}}, \bibinfo
  {author} {\bibfnamefont {D.}~\bibnamefont {Gu\'ery-Odelin}}, \ and\ \bibinfo
  {author} {\bibfnamefont {J.~G.}\ \bibnamefont {Muga}},\ }\bibfield  {title}
  {\enquote {\bibinfo {title} {Fast atomic transport without vibrational
  heating},}\ }\href {\doibase 10.1103/PhysRevA.83.013415} {\bibfield
  {journal} {\bibinfo  {journal} {Phys. Rev. A}\ }\textbf {\bibinfo {volume}
  {83}},\ \bibinfo {pages} {013415} (\bibinfo {year} {2011})}\BibitemShut
  {NoStop}%
\bibitem [{\citenamefont {Lewis}\ and\ \citenamefont
  {Riesenfeld}(1969)}]{Lewis1969}%
  \BibitemOpen
  \bibfield  {author} {\bibinfo {author} {\bibfnamefont {H.~R.}\ \bibnamefont
  {Lewis}}\ and\ \bibinfo {author} {\bibfnamefont {W.~B.}\ \bibnamefont
  {Riesenfeld}},\ }\bibfield  {title} {\enquote {\bibinfo {title} {An exact
  quantum theory of the time‐dependent harmonic oscillator and of a charged
  particle in a time‐dependent electromagnetic field},}\ }\href {\doibase
  10.1063/1.1664991} {\bibfield  {journal} {\bibinfo  {journal} {J. Math.
  Phys.}\ }\textbf {\bibinfo {volume} {10}},\ \bibinfo {pages} {1458--1473}
  (\bibinfo {year} {1969})}\BibitemShut {NoStop}%
\bibitem [{\citenamefont {Bradbury}\ \emph {et~al.}(2018)\citenamefont
  {Bradbury}, \citenamefont {Frostig}, \citenamefont {Hawkins}, \citenamefont
  {Johnson}, \citenamefont {Leary}, \citenamefont {Maclaurin},\ and\
  \citenamefont {Wanderman-Milne}}]{jax2018github}%
  \BibitemOpen
  \bibfield  {author} {\bibinfo {author} {\bibfnamefont {James}\ \bibnamefont
  {Bradbury}}, \bibinfo {author} {\bibfnamefont {Roy}\ \bibnamefont {Frostig}},
  \bibinfo {author} {\bibfnamefont {Peter}\ \bibnamefont {Hawkins}}, \bibinfo
  {author} {\bibfnamefont {Matthew~James}\ \bibnamefont {Johnson}}, \bibinfo
  {author} {\bibfnamefont {Chris}\ \bibnamefont {Leary}}, \bibinfo {author}
  {\bibfnamefont {Dougal}\ \bibnamefont {Maclaurin}}, \ and\ \bibinfo {author}
  {\bibfnamefont {Skye}\ \bibnamefont {Wanderman-Milne}},\ }\href
  {http://github.com/google/jax} {\enquote {\bibinfo {title} {{JAX}: composable
  transformations of {P}ython+{N}um{P}y programs},}\ } (\bibinfo {year}
  {2018})\BibitemShut {NoStop}%
\bibitem [{\citenamefont {Rumelhart}\ \emph {et~al.}(1986)\citenamefont
  {Rumelhart}, \citenamefont {Hinton},\ and\ \citenamefont
  {Williams}}]{Rumelhart1986}%
  \BibitemOpen
  \bibfield  {author} {\bibinfo {author} {\bibfnamefont {D.~E.}\ \bibnamefont
  {Rumelhart}}, \bibinfo {author} {\bibfnamefont {G.~E.}\ \bibnamefont
  {Hinton}}, \ and\ \bibinfo {author} {\bibfnamefont {R.~J.}\ \bibnamefont
  {Williams}},\ }\bibfield  {title} {\enquote {\bibinfo {title} {Learning
  representations by back-propagating errors},}\ }\href {\doibase
  10.1038/323533a0} {\bibfield  {journal} {\bibinfo  {journal} {Nature}\
  }\textbf {\bibinfo {volume} {323}},\ \bibinfo {pages} {533--536} (\bibinfo
  {year} {1986})}\BibitemShut {NoStop}%
\bibitem [{\citenamefont {Baur}\ and\ \citenamefont
  {Strassen}(1983)}]{BAUR1983317}%
  \BibitemOpen
  \bibfield  {author} {\bibinfo {author} {\bibfnamefont {Walter}\ \bibnamefont
  {Baur}}\ and\ \bibinfo {author} {\bibfnamefont {Volker}\ \bibnamefont
  {Strassen}},\ }\bibfield  {title} {\enquote {\bibinfo {title} {The complexity
  of partial derivatives},}\ }\href {\doibase
  https://doi.org/10.1016/0304-3975(83)90110-X} {\bibfield  {journal} {\bibinfo
   {journal} {Theoretical Computer Science}\ }\textbf {\bibinfo {volume}
  {22}},\ \bibinfo {pages} {317--330} (\bibinfo {year} {1983})}\BibitemShut
  {NoStop}%
\bibitem [{\citenamefont {Griewank}(1989)}]{Griewank89onautomatic}%
  \BibitemOpen
  \bibfield  {author} {\bibinfo {author} {\bibfnamefont {A}~\bibnamefont
  {Griewank}},\ }\bibfield  {title} {\enquote {\bibinfo {title} {On automatic
  differentiation},}\ }\href
  {https://citeseerx.ist.psu.edu/document?repid=rep1&type=pdf&doi=034212770af42af1db1018807398dbde1308a991}
  {\bibfield  {journal} {\bibinfo  {journal} {Mathematical Programming: Recent
  Developments and Applications}\ }\textbf {\bibinfo {volume} {6}},\ \bibinfo
  {pages} {83} (\bibinfo {year} {1989})}\BibitemShut {NoStop}%
\bibitem [{\citenamefont {Kingma}\ and\ \citenamefont {Ba}(2015)}]{adam}%
  \BibitemOpen
  \bibfield  {author} {\bibinfo {author} {\bibfnamefont {D.~P.}\ \bibnamefont
  {Kingma}}\ and\ \bibinfo {author} {\bibfnamefont {J.}~\bibnamefont {Ba}},\
  }\bibfield  {title} {\enquote {\bibinfo {title} {Adam: {A} method for
  stochastic optimization},}\ }\href {http://arxiv.org/abs/1412.6980}
  {\bibfield  {journal} {\bibinfo  {journal} {3rd International Conference on
  Learning Representations, {ICLR} 2015, San Diego, CA, USA, May 7-9, 2015,
  Conference Track Proceedings}\ } (\bibinfo {year} {2015})}\BibitemShut
  {NoStop}%
\bibitem [{\citenamefont {Porotti}\ \emph {et~al.}(2019)\citenamefont
  {Porotti}, \citenamefont {Tamascelli}, \citenamefont {Restelli},\ and\
  \citenamefont {Prati}}]{Tamascelli}%
  \BibitemOpen
  \bibfield  {author} {\bibinfo {author} {\bibfnamefont {R.}~\bibnamefont
  {Porotti}}, \bibinfo {author} {\bibfnamefont {D.}~\bibnamefont {Tamascelli}},
  \bibinfo {author} {\bibfnamefont {M}~\bibnamefont {Restelli}}, \ and\
  \bibinfo {author} {\bibfnamefont {E.}~\bibnamefont {Prati}},\ }\bibfield
  {title} {\enquote {\bibinfo {title} {Coherent transport of quantum states by
  deep reinforcement learning},}\ }\href {\doibase 10.1038/s42005-019-0169-x}
  {\bibfield  {journal} {\bibinfo  {journal} {Commun. Phys.}\ }\textbf
  {\bibinfo {volume} {2}},\ \bibinfo {pages} {61} (\bibinfo {year}
  {2019})}\BibitemShut {NoStop}%
\bibitem [{Note1()}]{Note1}%
  \BibitemOpen
  \bibinfo {note} {In our simulations we take $\Delta t\protect \tmspace
  -\thinmuskip {.1667em}=\protect \tmspace -\thinmuskip {.1667em}0.1$ which is
  sufficient to ensure good convergence.}\BibitemShut {Stop}%
\bibitem [{\citenamefont {Campbell}\ \emph {et~al.}(2015)\citenamefont
  {Campbell}, \citenamefont {De~Chiara}, \citenamefont {Paternostro},
  \citenamefont {Palma},\ and\ \citenamefont {Fazio}}]{CampbellPRL2015}%
  \BibitemOpen
  \bibfield  {author} {\bibinfo {author} {\bibfnamefont {S.}~\bibnamefont
  {Campbell}}, \bibinfo {author} {\bibfnamefont {G.}~\bibnamefont {De~Chiara}},
  \bibinfo {author} {\bibfnamefont {M.}~\bibnamefont {Paternostro}}, \bibinfo
  {author} {\bibfnamefont {G.~M.}\ \bibnamefont {Palma}}, \ and\ \bibinfo
  {author} {\bibfnamefont {R.}~\bibnamefont {Fazio}},\ }\bibfield  {title}
  {\enquote {\bibinfo {title} {Shortcut to adiabaticity in the
  lipkin-meshkov-glick model},}\ }\href {\doibase
  10.1103/PhysRevLett.114.177206} {\bibfield  {journal} {\bibinfo  {journal}
  {Phys. Rev. Lett.}\ }\textbf {\bibinfo {volume} {114}},\ \bibinfo {pages}
  {177206} (\bibinfo {year} {2015})}\BibitemShut {NoStop}%
\bibitem [{\citenamefont {Saberi}\ \emph {et~al.}(2014)\citenamefont {Saberi},
  \citenamefont {Opatrn\'y}, \citenamefont {M\o{}lmer},\ and\ \citenamefont
  {del Campo}}]{AdolfoPRA}%
  \BibitemOpen
  \bibfield  {author} {\bibinfo {author} {\bibfnamefont {H.}~\bibnamefont
  {Saberi}}, \bibinfo {author} {\bibfnamefont {T.}~\bibnamefont {Opatrn\'y}},
  \bibinfo {author} {\bibfnamefont {K.}~\bibnamefont {M\o{}lmer}}, \ and\
  \bibinfo {author} {\bibfnamefont {A.}~\bibnamefont {del Campo}},\ }\bibfield
  {title} {\enquote {\bibinfo {title} {Adiabatic tracking of quantum many-body
  dynamics},}\ }\href {\doibase 10.1103/PhysRevA.90.060301} {\bibfield
  {journal} {\bibinfo  {journal} {Phys. Rev. A}\ }\textbf {\bibinfo {volume}
  {90}},\ \bibinfo {pages} {060301(R)} (\bibinfo {year} {2014})}\BibitemShut
  {NoStop}%
\bibitem [{\citenamefont {Bartels}\ and\ \citenamefont
  {Mintert}(2013)}]{Referee3}%
  \BibitemOpen
  \bibfield  {author} {\bibinfo {author} {\bibfnamefont {Bj\"orn}\ \bibnamefont
  {Bartels}}\ and\ \bibinfo {author} {\bibfnamefont {Florian}\ \bibnamefont
  {Mintert}},\ }\bibfield  {title} {\enquote {\bibinfo {title} {Smooth optimal
  control with floquet theory},}\ }\href {\doibase 10.1103/PhysRevA.88.052315}
  {\bibfield  {journal} {\bibinfo  {journal} {Phys. Rev. A}\ }\textbf {\bibinfo
  {volume} {88}},\ \bibinfo {pages} {052315} (\bibinfo {year}
  {2013})}\BibitemShut {NoStop}%
\bibitem [{\citenamefont {Meister}\ \emph {et~al.}(2014)\citenamefont
  {Meister}, \citenamefont {Stockburger}, \citenamefont {Schmidt},\ and\
  \citenamefont {Ankerhold}}]{Referee4}%
  \BibitemOpen
  \bibfield  {author} {\bibinfo {author} {\bibfnamefont {Selina}\ \bibnamefont
  {Meister}}, \bibinfo {author} {\bibfnamefont {Jürgen~T}\ \bibnamefont
  {Stockburger}}, \bibinfo {author} {\bibfnamefont {Rebecca}\ \bibnamefont
  {Schmidt}}, \ and\ \bibinfo {author} {\bibfnamefont {Joachim}\ \bibnamefont
  {Ankerhold}},\ }\bibfield  {title} {\enquote {\bibinfo {title} {Optimal
  control theory with arbitrary superpositions of waveforms},}\ }\href
  {\doibase 10.1088/1751-8113/47/49/495002} {\bibfield  {journal} {\bibinfo
  {journal} {Journal of Physics A: Mathematical and Theoretical}\ }\textbf
  {\bibinfo {volume} {47}},\ \bibinfo {pages} {495002} (\bibinfo {year}
  {2014})}\BibitemShut {NoStop}%
\bibitem [{\citenamefont {Skinner}\ and\ \citenamefont
  {Gershenzon}(2010)}]{Referee5}%
  \BibitemOpen
  \bibfield  {author} {\bibinfo {author} {\bibfnamefont {Thomas~E.}\
  \bibnamefont {Skinner}}\ and\ \bibinfo {author} {\bibfnamefont {Naum~I.}\
  \bibnamefont {Gershenzon}},\ }\bibfield  {title} {\enquote {\bibinfo {title}
  {Optimal control design of pulse shapes as analytic functions},}\ }\href
  {\doibase 10.1016/j.jmr.2010.03.002} {\bibfield  {journal} {\bibinfo
  {journal} {Journal of Magnetic Resonance}\ }\textbf {\bibinfo {volume}
  {204}},\ \bibinfo {pages} {248--255} (\bibinfo {year} {2010})}\BibitemShut
  {NoStop}%
\bibitem [{\citenamefont {Wang}\ \emph {et~al.}(2016)\citenamefont {Wang},
  \citenamefont {Burgarth},\ and\ \citenamefont {Schirmer}}]{Wang2016}%
  \BibitemOpen
  \bibfield  {author} {\bibinfo {author} {\bibfnamefont {Xiaoting}\
  \bibnamefont {Wang}}, \bibinfo {author} {\bibfnamefont {Daniel}\ \bibnamefont
  {Burgarth}}, \ and\ \bibinfo {author} {\bibfnamefont {S.}~\bibnamefont
  {Schirmer}},\ }\bibfield  {title} {\enquote {\bibinfo {title} {Subspace
  controllability of spin-$\frac{1}{2}$ chains with symmetries},}\ }\href
  {\doibase 10.1103/PhysRevA.94.052319} {\bibfield  {journal} {\bibinfo
  {journal} {Phys. Rev. A}\ }\textbf {\bibinfo {volume} {94}},\ \bibinfo
  {pages} {052319} (\bibinfo {year} {2016})}\BibitemShut {NoStop}%
\bibitem [{\citenamefont {Lee}\ \emph {et~al.}(2018)\citenamefont {Lee},
  \citenamefont {Arenz}, \citenamefont {Rabitz},\ and\ \citenamefont
  {Russell}}]{Lee2018}%
  \BibitemOpen
  \bibfield  {author} {\bibinfo {author} {\bibfnamefont {Juneseo}\ \bibnamefont
  {Lee}}, \bibinfo {author} {\bibfnamefont {Christian}\ \bibnamefont {Arenz}},
  \bibinfo {author} {\bibfnamefont {Herschel}\ \bibnamefont {Rabitz}}, \ and\
  \bibinfo {author} {\bibfnamefont {Benjamin}\ \bibnamefont {Russell}},\
  }\bibfield  {title} {\enquote {\bibinfo {title} {Dependence of the quantum
  speed limit on system size and control complexity},}\ }\href {\doibase
  10.1088/1367-2630/aac6f3} {\bibfield  {journal} {\bibinfo  {journal} {New
  Journal of Physics}\ }\textbf {\bibinfo {volume} {20}},\ \bibinfo {pages}
  {063002} (\bibinfo {year} {2018})}\BibitemShut {NoStop}%
\bibitem [{\citenamefont {Goerz}\ \emph {et~al.}(2014)\citenamefont {Goerz},
  \citenamefont {Halperin}, \citenamefont {Aytac}, \citenamefont {Koch},\ and\
  \citenamefont {Whaley}}]{Referee1}%
  \BibitemOpen
  \bibfield  {author} {\bibinfo {author} {\bibfnamefont {Michael~H.}\
  \bibnamefont {Goerz}}, \bibinfo {author} {\bibfnamefont {Eli~J.}\
  \bibnamefont {Halperin}}, \bibinfo {author} {\bibfnamefont {Jon~M.}\
  \bibnamefont {Aytac}}, \bibinfo {author} {\bibfnamefont {Christiane~P.}\
  \bibnamefont {Koch}}, \ and\ \bibinfo {author} {\bibfnamefont {K.~Birgitta}\
  \bibnamefont {Whaley}},\ }\bibfield  {title} {\enquote {\bibinfo {title}
  {Robustness of high-fidelity rydberg gates with single-site
  addressability},}\ }\href {\doibase 10.1103/PhysRevA.90.032329} {\bibfield
  {journal} {\bibinfo  {journal} {Phys. Rev. A}\ }\textbf {\bibinfo {volume}
  {90}},\ \bibinfo {pages} {032329} (\bibinfo {year} {2014})}\BibitemShut
  {NoStop}%
\bibitem [{\citenamefont {Couvert}\ \emph {et~al.}(2008)\citenamefont
  {Couvert}, \citenamefont {Kawalec}, \citenamefont {Reinaudi},\ and\
  \citenamefont {Gu{\'{e}}ry-Odelin}}]{Couvert2008}%
  \BibitemOpen
  \bibfield  {author} {\bibinfo {author} {\bibfnamefont {A.}~\bibnamefont
  {Couvert}}, \bibinfo {author} {\bibfnamefont {T.}~\bibnamefont {Kawalec}},
  \bibinfo {author} {\bibfnamefont {G.}~\bibnamefont {Reinaudi}}, \ and\
  \bibinfo {author} {\bibfnamefont {D.}~\bibnamefont {Gu{\'{e}}ry-Odelin}},\
  }\bibfield  {title} {\enquote {\bibinfo {title} {Optimal transport of
  ultracold atoms in the non-adiabatic regime},}\ }\href {\doibase
  10.1209/0295-5075/83/13001} {\bibfield  {journal} {\bibinfo  {journal} {{EPL}
  (Europhysics Letters)}\ }\textbf {\bibinfo {volume} {83}},\ \bibinfo {pages}
  {13001} (\bibinfo {year} {2008})}\BibitemShut {NoStop}%
\bibitem [{\citenamefont {Larrouy}\ \emph {et~al.}(2020)\citenamefont
  {Larrouy}, \citenamefont {Patsch}, \citenamefont {Richaud}, \citenamefont
  {Raimond}, \citenamefont {Brune}, \citenamefont {Koch},\ and\ \citenamefont
  {Gleyzes}}]{PatschPRX}%
  \BibitemOpen
  \bibfield  {author} {\bibinfo {author} {\bibfnamefont {A.}~\bibnamefont
  {Larrouy}}, \bibinfo {author} {\bibfnamefont {S.}~\bibnamefont {Patsch}},
  \bibinfo {author} {\bibfnamefont {R.}~\bibnamefont {Richaud}}, \bibinfo
  {author} {\bibfnamefont {J.-M.}\ \bibnamefont {Raimond}}, \bibinfo {author}
  {\bibfnamefont {M.}~\bibnamefont {Brune}}, \bibinfo {author} {\bibfnamefont
  {C.~P.}\ \bibnamefont {Koch}}, \ and\ \bibinfo {author} {\bibfnamefont
  {S.}~\bibnamefont {Gleyzes}},\ }\bibfield  {title} {\enquote {\bibinfo
  {title} {Fast navigation in a large hilbert space using quantum optimal
  control},}\ }\href {\doibase 10.1103/PhysRevX.10.021058} {\bibfield
  {journal} {\bibinfo  {journal} {Phys. Rev. X}\ }\textbf {\bibinfo {volume}
  {10}},\ \bibinfo {pages} {021058} (\bibinfo {year} {2020})}\BibitemShut
  {NoStop}%
\bibitem [{\citenamefont {Epstein}\ and\ \citenamefont
  {Whaley}(2017)}]{Epstein_2017}%
  \BibitemOpen
  \bibfield  {author} {\bibinfo {author} {\bibfnamefont {Jeffrey~M.}\
  \bibnamefont {Epstein}}\ and\ \bibinfo {author} {\bibfnamefont {K.~Birgitta}\
  \bibnamefont {Whaley}},\ }\bibfield  {title} {\enquote {\bibinfo {title}
  {Quantum speed limits for quantum-information-processing tasks},}\ }\href
  {\doibase 10.1103/PhysRevA.95.042314} {\bibfield  {journal} {\bibinfo
  {journal} {Phys. Rev. A}\ }\textbf {\bibinfo {volume} {95}},\ \bibinfo
  {pages} {042314} (\bibinfo {year} {2017})}\BibitemShut {NoStop}%
\bibitem [{\citenamefont {Murphy}\ \emph {et~al.}(2010)\citenamefont {Murphy},
  \citenamefont {Montangero}, \citenamefont {Giovannetti},\ and\ \citenamefont
  {Calarco}}]{Murphy2010}%
  \BibitemOpen
  \bibfield  {author} {\bibinfo {author} {\bibfnamefont {M.}~\bibnamefont
  {Murphy}}, \bibinfo {author} {\bibfnamefont {S.}~\bibnamefont {Montangero}},
  \bibinfo {author} {\bibfnamefont {V.}~\bibnamefont {Giovannetti}}, \ and\
  \bibinfo {author} {\bibfnamefont {T.}~\bibnamefont {Calarco}},\ }\bibfield
  {title} {\enquote {\bibinfo {title} {Communication at the quantum speed limit
  along a spin chain},}\ }\href {\doibase 10.1103/PhysRevA.82.022318}
  {\bibfield  {journal} {\bibinfo  {journal} {Phys. Rev. A}\ }\textbf {\bibinfo
  {volume} {82}},\ \bibinfo {pages} {022318} (\bibinfo {year}
  {2010})}\BibitemShut {NoStop}%
\bibitem [{\citenamefont {Krotov}(1993)}]{krotovGlobalMethodsOptimal1993}%
  \BibitemOpen
  \bibfield  {author} {\bibinfo {author} {\bibfnamefont {V.~F.}\ \bibnamefont
  {Krotov}},\ }\bibfield  {title} {\enquote {\bibinfo {title} {Global
  {{Methods}} in {{Optimal Control Theory}}},}\ }in\ \href {\doibase
  10.1007/978-1-4612-0349-0_3} {\emph {\bibinfo {booktitle} {Advances in
  {{Nonlinear Dynamics}} and {{Control}}: {{A Report}} from {{Russia}}}}},\
  \bibinfo {series and number} {Progress in {{Systems}} and {{Control
  Theory}}},\ \bibinfo {editor} {edited by\ \bibinfo {editor} {\bibfnamefont
  {Alexander~B.}\ \bibnamefont {Kurzhanski}}}\ (\bibinfo  {publisher}
  {{Birkh\"auser}},\ \bibinfo {address} {{Boston, MA}},\ \bibinfo {year}
  {1993})\ pp.\ \bibinfo {pages} {74--121}\BibitemShut {NoStop}%
\bibitem [{\citenamefont {Burgarth}\ and\ \citenamefont
  {Bose}(2005)}]{Burgarth_2005}%
  \BibitemOpen
  \bibfield  {author} {\bibinfo {author} {\bibfnamefont {D.}~\bibnamefont
  {Burgarth}}\ and\ \bibinfo {author} {\bibfnamefont {S.}~\bibnamefont
  {Bose}},\ }\bibfield  {title} {\enquote {\bibinfo {title} {Perfect quantum
  state transfer with randomly coupled quantum chains},}\ }\href {\doibase
  10.1088/1367-2630/7/1/135} {\bibfield  {journal} {\bibinfo  {journal} {New J.
  Phys.}\ }\textbf {\bibinfo {volume} {7}},\ \bibinfo {pages} {135--135}
  (\bibinfo {year} {2005})}\BibitemShut {NoStop}%
\bibitem [{\citenamefont {De~Chiara}\ \emph {et~al.}(2005)\citenamefont
  {De~Chiara}, \citenamefont {Rossini}, \citenamefont {Montangero},\ and\
  \citenamefont {Fazio}}]{DeChiara2005}%
  \BibitemOpen
  \bibfield  {author} {\bibinfo {author} {\bibfnamefont {G.}~\bibnamefont
  {De~Chiara}}, \bibinfo {author} {\bibfnamefont {D.}~\bibnamefont {Rossini}},
  \bibinfo {author} {\bibfnamefont {S.}~\bibnamefont {Montangero}}, \ and\
  \bibinfo {author} {\bibfnamefont {R.}~\bibnamefont {Fazio}},\ }\bibfield
  {title} {\enquote {\bibinfo {title} {From perfect to fractal transmission in
  spin chains},}\ }\href {\doibase 10.1103/PhysRevA.72.012323} {\bibfield
  {journal} {\bibinfo  {journal} {Phys. Rev. A}\ }\textbf {\bibinfo {volume}
  {72}},\ \bibinfo {pages} {012323} (\bibinfo {year} {2005})}\BibitemShut
  {NoStop}%
\bibitem [{\citenamefont {Mishra}\ \emph {et~al.}(2021)\citenamefont {Mishra},
  \citenamefont {Trivedi}, \citenamefont {Safavi-Naeini},\ and\ \citenamefont
  {Vu\ifmmode \check{c}\else \v{c}\fi{}kovi\ifmmode~\acute{c}\else
  \'{c}\fi{}}}]{Referee2}%
  \BibitemOpen
  \bibfield  {author} {\bibinfo {author} {\bibfnamefont {Sattwik~Deb}\
  \bibnamefont {Mishra}}, \bibinfo {author} {\bibfnamefont {Rahul}\
  \bibnamefont {Trivedi}}, \bibinfo {author} {\bibfnamefont {Amir~H.}\
  \bibnamefont {Safavi-Naeini}}, \ and\ \bibinfo {author} {\bibfnamefont
  {Jelena}\ \bibnamefont {Vu\ifmmode \check{c}\else
  \v{c}\fi{}kovi\ifmmode~\acute{c}\else \'{c}\fi{}}},\ }\bibfield  {title}
  {\enquote {\bibinfo {title} {Control design for inhomogeneous-broadening
  compensation in single-photon transducers},}\ }\href {\doibase
  10.1103/PhysRevApplied.16.044025} {\bibfield  {journal} {\bibinfo  {journal}
  {Phys. Rev. Applied}\ }\textbf {\bibinfo {volume} {16}},\ \bibinfo {pages}
  {044025} (\bibinfo {year} {2021})}\BibitemShut {NoStop}%
\bibitem [{\citenamefont {Sutton}\ and\ \citenamefont
  {Barto}(2018)}]{Sutton2018}%
  \BibitemOpen
  \bibfield  {author} {\bibinfo {author} {\bibfnamefont {Richard~S.}\
  \bibnamefont {Sutton}}\ and\ \bibinfo {author} {\bibfnamefont {Andrew~G.}\
  \bibnamefont {Barto}},\ }\href@noop {} {\emph {\bibinfo {title}
  {Reinforcement Learning: An Introduction}}}\ (\bibinfo  {publisher} {A
  Bradford Book},\ \bibinfo {address} {Cambridge, MA, USA},\ \bibinfo {year}
  {2018})\BibitemShut {NoStop}%
\bibitem [{\citenamefont {Zhang}\ \emph {et~al.}(2019)\citenamefont {Zhang},
  \citenamefont {Wei}, \citenamefont {Asad}, \citenamefont {Yang},\ and\
  \citenamefont {Wang}}]{zhangWhenDoesReinforcement2019}%
  \BibitemOpen
  \bibfield  {author} {\bibinfo {author} {\bibfnamefont {Xiao-Ming}\
  \bibnamefont {Zhang}}, \bibinfo {author} {\bibfnamefont {Zezhu}\ \bibnamefont
  {Wei}}, \bibinfo {author} {\bibfnamefont {Raza}\ \bibnamefont {Asad}},
  \bibinfo {author} {\bibfnamefont {Xu-Chen}\ \bibnamefont {Yang}}, \ and\
  \bibinfo {author} {\bibfnamefont {Xin}\ \bibnamefont {Wang}},\ }\bibfield
  {title} {\enquote {\bibinfo {title} {When does reinforcement learning stand
  out in quantum control? {{A}} comparative study on state preparation},}\
  }\href {\doibase 10.1038/s41534-019-0201-8} {\bibfield  {journal} {\bibinfo
  {journal} {npj Quantum Information}\ }\textbf {\bibinfo {volume} {5}},\
  \bibinfo {pages} {1--7} (\bibinfo {year} {2019})}\BibitemShut {NoStop}%
\bibitem [{\citenamefont {Erdman}\ and\ \citenamefont
  {No{\'e}}(2022)}]{Erdman2021}%
  \BibitemOpen
  \bibfield  {author} {\bibinfo {author} {\bibfnamefont {Paolo~A.}\
  \bibnamefont {Erdman}}\ and\ \bibinfo {author} {\bibfnamefont {Frank}\
  \bibnamefont {No{\'e}}},\ }\bibfield  {title} {\enquote {\bibinfo {title}
  {Identifying optimal cycles in quantum thermal machines with
  reinforcement-learning},}\ }\href {\doibase 10.1038/s41534-021-00512-0}
  {\bibfield  {journal} {\bibinfo  {journal} {npj Quantum Information}\
  }\textbf {\bibinfo {volume} {8}},\ \bibinfo {pages} {1} (\bibinfo {year}
  {2022})}\BibitemShut {NoStop}%
\bibitem [{\citenamefont {Erdman}\ and\ \citenamefont
  {Noé}(2022)}]{Erdman2022}%
  \BibitemOpen
  \bibfield  {author} {\bibinfo {author} {\bibfnamefont {Paolo~Andrea}\
  \bibnamefont {Erdman}}\ and\ \bibinfo {author} {\bibfnamefont {Frank}\
  \bibnamefont {Noé}},\ }\href {\doibase 10.48550/ARXIV.2204.04785} {\enquote
  {\bibinfo {title} {Driving black-box quantum thermal machines with optimal
  power/efficiency trade-offs using reinforcement learning},}\ } (\bibinfo
  {year} {2022})\BibitemShut {NoStop}%
\bibitem [{\citenamefont {Wierstra}\ \emph {et~al.}(2014)\citenamefont
  {Wierstra}, \citenamefont {Schaul}, \citenamefont {Glasmachers},
  \citenamefont {Sun}, \citenamefont {Peters},\ and\ \citenamefont
  {Schmidhuber}}]{JMLR:v15:wierstra14a}%
  \BibitemOpen
  \bibfield  {author} {\bibinfo {author} {\bibfnamefont {Daan}\ \bibnamefont
  {Wierstra}}, \bibinfo {author} {\bibfnamefont {Tom}\ \bibnamefont {Schaul}},
  \bibinfo {author} {\bibfnamefont {Tobias}\ \bibnamefont {Glasmachers}},
  \bibinfo {author} {\bibfnamefont {Yi}~\bibnamefont {Sun}}, \bibinfo {author}
  {\bibfnamefont {Jan}\ \bibnamefont {Peters}}, \ and\ \bibinfo {author}
  {\bibfnamefont {J{\"u}rgen}\ \bibnamefont {Schmidhuber}},\ }\bibfield
  {title} {\enquote {\bibinfo {title} {Natural evolution strategies},}\
  }\href@noop {} {\bibfield  {journal} {\bibinfo  {journal} {Journal of Machine
  Learning Research}\ }\textbf {\bibinfo {volume} {15}},\ \bibinfo {pages}
  {949--980} (\bibinfo {year} {2014})}\BibitemShut {NoStop}%
\bibitem [{\citenamefont {Day}\ \emph {et~al.}(2019)\citenamefont {Day},
  \citenamefont {Bukov}, \citenamefont {Weinberg}, \citenamefont {Mehta},\ and\
  \citenamefont {Sels}}]{Day_2019}%
  \BibitemOpen
  \bibfield  {author} {\bibinfo {author} {\bibfnamefont {Alexandre G.~R.}\
  \bibnamefont {Day}}, \bibinfo {author} {\bibfnamefont {Marin}\ \bibnamefont
  {Bukov}}, \bibinfo {author} {\bibfnamefont {Phillip}\ \bibnamefont
  {Weinberg}}, \bibinfo {author} {\bibfnamefont {Pankaj}\ \bibnamefont
  {Mehta}}, \ and\ \bibinfo {author} {\bibfnamefont {Dries}\ \bibnamefont
  {Sels}},\ }\bibfield  {title} {\enquote {\bibinfo {title} {Glassy phase of
  optimal quantum control},}\ }\href {\doibase 10.1103/PhysRevLett.122.020601}
  {\bibfield  {journal} {\bibinfo  {journal} {Phys. Rev. Lett.}\ }\textbf
  {\bibinfo {volume} {122}},\ \bibinfo {pages} {020601} (\bibinfo {year}
  {2019})}\BibitemShut {NoStop}%
\bibitem [{\citenamefont {Dalgaard}\ \emph {et~al.}(2022)\citenamefont
  {Dalgaard}, \citenamefont {Motzoi},\ and\ \citenamefont
  {Sherson}}]{dalgaard2021predicting}%
  \BibitemOpen
  \bibfield  {author} {\bibinfo {author} {\bibfnamefont {Mogens}\ \bibnamefont
  {Dalgaard}}, \bibinfo {author} {\bibfnamefont {Felix}\ \bibnamefont
  {Motzoi}}, \ and\ \bibinfo {author} {\bibfnamefont {Jacob}\ \bibnamefont
  {Sherson}},\ }\bibfield  {title} {\enquote {\bibinfo {title} {Predicting
  quantum dynamical cost landscapes with deep learning},}\ }\href {\doibase
  10.1103/PhysRevA.105.012402} {\bibfield  {journal} {\bibinfo  {journal}
  {Phys. Rev. A}\ }\textbf {\bibinfo {volume} {105}},\ \bibinfo {pages}
  {012402} (\bibinfo {year} {2022})}\BibitemShut {NoStop}%
\end{thebibliography}%

\end{document}